\documentclass{aastex63}
\usepackage{bm}
\usepackage{CJKutf8}
\usepackage{amsmath}
\usepackage{comment}

\pdfoutput=1

\graphicspath{{./}{figures/}}

\shorttitle{QPPs in Magnetic Reconnection Rate}
\shortauthors{Corchado-Albelo et al.}

\begin{document}

\title{Inferring Fundamental Properties of the Flare Current Sheet Using Flare Ribbons:\\Oscillations in the Reconnection Flux Rates}

\correspondingauthor{Marcel F. Corchado-Albelo}
\email{marcel.corchado@colorado.edu}

\author[0000-0003-1597-0184]{Marcel~F.~Corchado Albelo}
\altaffiliation{DKIST Ambassador}
\affil{Department of Astrophysical and Planetary Sciences, University of Colorado Boulder, 2000 Colorado Avenue, Boulder, CO 80305, USA}
\affiliation{National Solar Observatory, 3665 Discovery Drive, Boulder, CO 80303, USA}

\author[0000-0001-8975-7605]{Maria~D.~Kazachenko}
\affiliation{Department of Astrophysical and Planetary Sciences, University of Colorado Boulder, 2000 Colorado Avenue, Boulder, CO 80305, USA}
\affiliation{National Solar Observatory, 3665 Discovery Drive, Boulder, CO 80303, USA}

\author[0000-0001-6886-855X]{Benjamin~J.~Lynch}
\affiliation{Space Sciences Laboratory, University of California, Berkeley, CA 94720, USA}
\affiliation{Department of Earth, Planetary, and Space Sciences, University of California, Los Angeles, CA 90095, USA}

\begin{abstract}
Magnetic reconnection is understood to be the main physical process that facilitates the transformation of magnetic energy into heat, motion, and particle acceleration during solar eruptions. Yet, observational constraints on  reconnection region properties and dynamics are limited due to lack of high-cadence and high-spatial-resolution observations. By studying the evolution and morphology of post-reconnected field-lines footpoints, or flare ribbons and vector photospheric magnetic field, we estimate the magnetic reconnection flux and its rate of change with time to study the flare reconnection process and dynamics of the current sheet above. We compare high-resolution imaging data to study the evolution of the fine structure in flare ribbons as ribbons spread away from the polarity inversion line. Using data from two illustrative events (one M- and X-class flare), we explore the relationship between the ribbon-front fine structure and the temporal development of bursts in the reconnection region. Additionally, we use the \verb+RibbonDB+ database to perform statistical analysis of 73 (C- to X-class) flares and identify QPP’s properties using the Wavelet Transform. Our main finding is the discovery of quasi-periodic pulsations (QPP) signatures in the derived magnetic reconnection rates in both example events and the large flare sample. We find that the oscillations’ periods range from one to four minutes. Furthermore, we find nearly co-temporal bursts in Hard X-ray (HXR) emission profiles. We discuss how dynamical processes in the current sheet involving plasmoids can explain the nearly-co-temporal signatures of quasi periodicity in the reconnection rates and HXR emission.

\end{abstract}
%%%%%%%%%%%%%%%%%%%%%%%%%%%%%%%%%%%%%%%%%%%%%%%%

\section{Introduction}\label{sec:intr}
Solar flares are intense localized emission of light in the solar atmosphere that cover the full electromagnetic spectrum  from radio to $\gamma$-rays (see review by \citealt{Benz2017FlareObservations}). They are understood to be associated with the release of free magnetic energy stored in the twisted and sheared coronal magnetic fields via magnetic reconnection (see reviews by \citealt{Priest2002TheFlares,Hudson2011GlobalFlares,Shibata2011SolarProcesses}). Although there is no direct measurement of the coronal magnetic reconnection, indirect observations and models have helped to gain understanding of this process.

The CSHKP or a standard two-ribbon flare model named after \citet{Carmichael1964AFlares,Sturrock1966ModelFlares,Hirayama1974TheoreticalProminences,Kopp1976MagneticPhenomenon} explains many features in observations and numerical experiments. For example, it describes a sequence of coronal (flare) loops formed by reconnection as the magnetic reconnection point (so called X-point in 2D) moves upwards beneath an erupting flux rope (EFR; O-point in 2D) that is a CME. These loops are anchored in opposite magnetic polarities, and the line separating these is referred to as the polarity inversion line (PIL). Non-thermal particles accelerated directly or indirectly during the reconnection process precipitate down the newly formed flare loops and arrive to the chromosphere. At the chromosphere the density is high enough for the accelerated electrons and ions to deposit energy through bremsstrahlung and produce electromagnetic radiation at various wavelengths (e.g. H$_\alpha$, ultraviolet, hard X-rays, $\gamma$-rays; \citealt{Brown1971TheBursts}). As the new flaring loops move higher into the corona, the location of the energy deposition moves away from the PIL. As the chromosphere is locally heated by condensed non-thermal particles, thermal conduction forces chromospheric material to rise (so called chromospheric evaporation) and fill the newly reconnected flare loops, generating emission in soft X-rays (SXR) within the flare arcade. Understanding of the observed phenomena described above critically depends on our understanding of the details of magnetic reconnection.  

Magnetic reconnection can proceed through different mechanisms which will dictate the rate through which new field lines are reconfigured. The simplest magnetic reconnection mechanism to explain the speed at which anti-parallel magnetic field lines enter the reconnection site and have their magnetic topologies changed is the Sweet-Parker (SP) mechanism \citep{Sweet1958,Sweet1958b,Parker1957SweetsFluids,Parker1963TheFields.}. This mechanism treats a single laminar layer current sheet stretching along the full interface between the opposing magnetic fields. The interface region is surrounded by plasma in ideal magnetohydrodynamic (MHD) conditions. {This leads to a slow reconnection rate, or inflow speed of new field lines to the reconnection site, of field lines  $M_A =\tfrac{V_R}{V_A} = S^{-1/2}$, where $S = L V_A/ \eta$ is the Lundquist number, $\eta$ is the plasma resistivity, $L$ the characteristic length of of the system, and $V_A$ is the Alfv\'{e}n speed. This dependence of the Lundquist number yields a reconnection rate that is too slow to explain the energy release time during solar flares.} The required inflow velocity for solar flares is on the order of tenths or hundredths of the Alfv\'{e}n speed and is called fast reconnection. {This requires that the reconnection rate has a weaker scaling dependence with the Lundquist number \citep{Pucci2017FastEffect}.} One alternative mechanism to explain fast reconnection is the Petschek model \citep{Petschek1964}. The Petschek model  assumes a smaller reconnection site, compared to the current sheet length. This geometry creates a larger opening on the reconnection exhaust, which increases the outflow velocity of the newly formed magnetic fields. Another mechanism for fast reconnection is the Hall MHD, in which the current sheet is required to have a  width of the ion inertial length. Lastly, another possibility of fast reconnection is through impulsive or bursty reconnection. Such a process requires an elongated SP-like current sheet undergoing a tearing instability. For a detailed review of magnetic reconnection mechanisms and their application to solar flares see \citet{Pontin2022MagneticModelling} and {\citet{Ji2022MagneticExperiments}}.    

{The tearing instability (TI) is when the CS becomes unstable and tears, breaks, and subsequently forms magnetic island plasmoids within the current layer \citep{Furth1963Finite-resistivityPinch, Shibata2001Plasmoid-induced-reconnectionReconnection}}.  \cite{Shibata2001Plasmoid-induced-reconnectionReconnection} showed a scenario of cascading reconnection that results in bursty reconnection episodes. In this scenario initially the CS is stretched out, then the TI mode sets and fragments the CS, which form multiple plasmoids. The plasmoids separate as they outflow towards the CS exhausts and trigger secondary and higher levels of TI{, which we refer to as the plasmoid instability (PI) in order to account for the interactions, and dynamics of the plasmoid structures (i.e., the effect of the TI on the further evolution of the magnetic reconnection process in the system).} The higher level TI can continue forming plasmoids of smaller sizes until reaching the plasma-kinetic scale. This scenario has been supported both analytically \citep{Loureiro2007InstabilityChains,Uzdensky2010FastRegime} and with numerical simulations (e.g., \citealt{Forbes.Isenberg1991_rcc,Karpen1998,Karpen2012,Bhattacharjee2009FastInstability,Lin2009,Samtaney2009,Huang2010ScalingRegime,Barta2011SPONTANEOUSANALYSIS, Shen2011NUMERICALFLARES,Mei2012NumericalSheets,Ni2015FASTINSTABILITY,Huang2016TURBULENTINSTABILITY,Lynch2016,Mei2017MagneticRopes,Li2019FormationReconnection,Wang2022Current-sheetFlares}). Yet, with no clear observations of the coronal magnetic field in the reconnection site, and limited temporal and spatial cadence it is hard to directly observe these structures and their dynamics in the flare current sheets.

There is observational evidence showing blob like structures propagating within current sheets in EUV, SXR, and radio observations, which have been interpreted as plasmoid structures within the current sheet (e.g., \citealt{Ohyama1998XRayEjection, Kliem2000SolarReconnection, Takasao2012SIMULTANEOUSFLARE,Hayes2019,Lu2022ObservationalFlare}). These observations support the existence of plasmoids in the flare current sheet, and that the TI and PI are present during flares. The plasmoids are usually observed as enhanced emission blobs or dark voids (depending on the energy range), propagating downwards and upwards along the current sheet. Studying these structures has allowed estimation of the reconnection properties of the CS like the outflow speed of magnetic reconnection \citep{Takasao2012SIMULTANEOUSFLARE,Lu2022ObservationalFlare}, and provide links to quasi periodic pulsations in flare light curves \citep{Hayes2019}.

Flare ribbons are localized enhancements of emission in the upper chromosphere (some 2000 km above the solar surface). The flare ribbons are usually observed in H$_\alpha$ and $1600$ \AA{} ultraviolet (UV) emission. They are interpreted as the signature of non-thermal particles precipitating from a coronal source, due to magnetic reconnection, into the dense chromosphere marking the footpoints of newly reconnected flare loops (\citealt{Forbes2000,Fletcher2011,Qiu2012,Longcope2014,Li2014,Li2017,Graham2015,Priest2017}). As so, they provide a powerful tool to indirectly observe and diagnose the current sheet properties during magnetic reconnection.

\cite{Forbes2000WhatEjections,Forbes.Priest1984_rcc} described the quantitative relationship between the reconnected flux rate and the motion of the flare ribbons within the CSHKP model. They used the rate of photospheric magnetic flux change $\dot{\Phi}_\mathrm{phot}$ swept by the flare ribbons to determine {\it the magnetic reconnection rate}, i.e. the rate at which the coronal magnetic flux is processed through current sheet in 3D:  
\begin{equation}
	\frac{\partial \Phi_\mathrm{c}}{\partial t} = \frac{\partial}{\partial t} \int B_\mathrm{c} dS_\mathrm{c} = \frac{\partial}{\partial t} \int B_\mathrm{n} dS_\mathrm{rbn}  = \frac{\partial \Phi_\mathrm{phot}}{\partial t}.
\end{equation}
Here the coronal magnetic field $B_\mathrm{c}$ reconnection rate $\frac{\partial \Phi_\mathrm{c}}{\partial t}$ is defined by the integration of an inflow coronal magnetic field over the reconnection area $dS_\mathrm{c}$. Unfortunately, this quantity is not readily available from observations due to difficulties in measuring the coronal magnetic field and the resolution needed to resolve the reconnection site area. However by conservation of flux we can use the change per unit time of flare ribbon magnetic flux $\frac{\partial \Phi_\mathrm{phot}}{\partial t}$ to infer the change of the coronal field during magnetic reconnection. In the expression above the magnetic flux $\Phi$ or $\Phi_\mathrm{phot} = \int B_\mathrm{n}dS_\mathrm{rbn}$, $B_\mathrm{n}$ is the normal component of the photospheric magnetic field and $dS_\mathrm{rbn}$ is the area swept by the flare ribbon. Measurements of the flare ribbon magnetic flux are relatively straightforward with current observation capabilities. This allows the creation of databases like the \verb+RibbonDB+ \citep{Kazachenko2017} which contains reconnection fluxes and other related properties for more than $3000$ flares.

High-resolution observations of swirls and wave-breaks in flare ribbons and the ribbon fronts (newly brightened kernels in the chromosphere) suggest that flare ribbons' evolution is connected to current sheet processes. Particularly, the flare ribbon structure has been linked to wave-generating processes at the flare loop tops \citep{Brannon2015SPECTROSCOPICWAVES} and to the existence of plasmoid structures within the current sheet (see \citealt{Wyper2021IsSheet,Naus2022CorrelatedFlare} and \S2.2 of the review by \citealt{Kazachenko2022r}). Therefore high-cadence and high-spatial-resolution studies of flare ribbons and their comparison with numerical flare models could potentially tell us about the physics of the current sheet and magnetic reconnection during flares, providing the missing link between the reconnection properties and other flare phenomena like particle acceleration. Particle-in-cell (PIC) models have shown that contracting magnetic islands can also accelerate particles trapped within them \citep{Drake2006ElectronReconnection,Guidoni2016MAGNETIC-ISLANDFLARES}. Recent theoretical work has shown that non-thermal particle populations traveling through consecutive magnetic islands can give rise to a population of particles capable of producing the typical SXR and hard X-ray (HXR) emission observed during flares \citep{Guidoni2022SpectralIslands}. They also provide a possible mechanism to explain particle acceleration contained within confined temporal enhancements or bursts, which might show periodic, oscillatory, or intermittent behavior.

Observations of solar and stellar flare light curves in multiple frequencies have shown oscillatory contributions usually referred to as Quasi Periodic Pulsation (QPPs; see reviews by \citealt{Nakariakov2009Quasi-PeriodicFlares,VanDoorsselaere2016Quasi-periodicReview}). Their existence within the whole spectrum of flare emission suggest that multiple mechanisms might be causing oscillatory and bursty particle acceleration (see review by \citealt{McLaughlin2018ModellingFlares}) including MHD waves \citep{Thurgood2017}, plasmoid interactions with the flaring loops and among themselves \citep{Kliem2000SolarReconnection,Barta2008PlasmoidStructure,Jelinek2017OscillationsSheet}, and existence of termination shocks due to the superposition of reconnection outflow jets \citep{Takahashi2017Quasi-periodicReconnection}. Therefore, observational constraints could help our understanding of what is the mechanism behind these QPPs which in turn could improve our modeling capabilities of solar and stellar flares.

In this paper, we report the observations of oscillation in the magnetic reconnection rates of dozens of flares. To our knowledge it is the largest survey up to date. Specifically, we present statistical analysis of $73$ events from the \verb+RibbonDB+ database which have co-temporal HXR and SXR observations. From our analysis, we provide observational reference relating oscillations in reconnection rates with particle acceleration proxies in HXR and SXR with the goal of understanding the details of the flare reconnection process and its relationship with particle acceleration.  The structure of the paper is as follows. In Section \ref{sec:met} we describe the data we used in this study, the methods used to derive the magnetic reconnection properties from flare ribbon and magnetogram observations, and the detection algorithm we used to identify the QPPs. In section \ref{sec:res} we first show examples of individual flares comparing spatial flare ribbon evolution to the evolution of the reconnection budget and the rate of its change with time; we then compare oscillating reconnection flux/rate evolution with co-temporal SXR and HXR emissions and summarize the results of the 73 events. In section \ref{sec:dis} we discuss and interpret our results in the context of recent works in the field. Finally, we summarize our findings in section \ref{sec:con}.

%%%%%%%%%%%%%%%%%%%%%%%%%%%%%%%%%%%%%%%%%%%%%%
\section{Data \& Methods}\label{sec:met}
\subsection{Data Description}
In this section we describe the observations from the {Helioseismic Magnetic Imager} (HMI) and {Atmospheric Image Assembly} (AIA) onboard the {Solar Dynamics Observatory} (SDO), {Interface Region Imaging Spectrograph} (IRIS), {Fermi Gamma-ray Space Telescope} (Fermi), and {Geostationary Operational Environmental Satellite} (GOES), the methodology for evaluating magnetic reconnection flux and and its rate and our analysis of the QPPs using the Wavelet Transform.

\subsubsection{SDO: AIA and HMI}
We use the data from SDO (\citealt{Pesnell2012TheSDO}) to derive flare ribbons' spatial properties and reconnection fluxes/rates. SDO provides the capability to derive high-quality spatial and temporal observations of the full-disk intensity maps in different narrow-band filters, and vector magnetic fields on the same platform, allowing the co-spatial observations of flare ribbons and their host active region's photospheric magnetic fields. AIA (\citealt{Lemen2012TheSDO}) observes in seven extreme-ultraviolet (EUV) and three UV channels. For the purpose of this study we use the $1600$ \AA{} intensity maps, corresponding to a characteristic temperature response of $10^5$ K and $5000$ K, sensitive to chromospheric emissions. The spatial resolution is $0.6$", and  cadence is $24$ seconds. HMI provides the photospheric magnetic field strength, inclination and azimuth \citep{Hoeksema2014ThePerformance}, which can be transformed into B$_x$, B$_y$, and B$_z$  \citep{Sun2013OnNote,Bobra2021Mbobra/SHARPs:2021-07-23}. To improve the quality of the signal, vector-field frames produced every $135$ s were combined into a $720$ second cadence observable, with a pixel resolution of $0.5$".

\subsubsection{IRIS: SJI Images}
For inferring the fine structure of flare ribbons we use observations from IRIS (\citealt{DePontieu2014TheIRIS,DePontieu2021AIRIS}). IRIS observes in Far Ultraviolet (FUV) and Near Ultraviolet (NUV) passbands since June 2013. The satellite has two science instruments, a Spectrograph (SG) and a Slit-Jaw Imager (SJI). The SG has a field of view (FOV) up to $175$", and spatial resolution of $0.33$" - $0.4$" for FUV and NUV respectively. It probes a wide temperature regime from the photosphere ($5000$ K) to the corona ($10^6$ K  - $10^7$ K), looking at FUV ($1332 $\AA{} - $1358$ \AA{} and  $1389$ \AA{} - $1407$ \AA{}), and NUV ($2783$ \AA{} - $2835$ \AA{}) lines. The SJI observes with a ranging cadence on the order of seconds with a spatial resolution of usually $0.166$" covering a FOV of up to $175$" $\times 175$", in four passbands probing  the transition region (C II $1335$ \AA{} \& Si IV $1400$ \AA{}), chromosphere (Mg II k $2796$ \AA{}), and photosphere (continuum $2830$ \AA{}). To extract and use the IRIS level 2 data, which have been corrected for flat field, dark currents, systematic offsets, and geometric correction with the IDL script \textit{iris\_prep.pro} \citep{DePontieu2021AIRIS}, we use the \textit{sunkit-instruments} package from the \textit{SunPy} Python library \citep{Barnes2020ThePackage}.

\subsubsection{Fermi: GBM}
To describe high-energy flare properties, we use observations from Fermi (\citealt{Atwood2009TheMission}). Fermi, has two instruments, the \textit{Large Area Telescope} (LAT) and the \textit{Gamma-ray Burst Monitor} (GBM, \citealt{Meegan2009TheMonitor}). These instruments are used to observe high energy astrophysical phenomena: high-energy $\gamma$-ray blazars, pulsars, $\gamma$-ray bursts, high-energy solar flares, etc. The LAT, serving as the primary instrument aboard Fermi, is a wide field-of-view imager covering an energy range of $\approx 20$ MeV to $\approx 300$ GeV. The GBM expands the LAT capabilities to energy ranges below those observed by LAT into the HXR range. The GBM produces three data types, out of which two are temporally binned, the Continuous Time (CTIME) and Continuous Spectroscopy (CSPEC), and the Time-Tagged Events (TTE). For large solar flares TTE data is usually lost, and thus CTIME and CSPEC are the best data products used to evaluate temporal and spectral evolution during flares. We use the \textit{OSPEX} IDL software\footnote{\url{https://hesperia.gsfc.nasa.gov/fermi\_solar/analyzing\_fermi\_gbm.htm}} to download CSPEC data, which has a varying cadence from $1.0$ seconds to $32.7$ seconds, and default value of 4.0 seconds stored in 128 quasi-logarithmic energy bins, which can be integrated into the following energy channels of interest: $8-15$ keV, $15-25$ keV, $25-50$ keV, $50-100$ keV, and $100-300$ keV.

\subsubsection{GOES: XRS}
Finally, to describe flare lightcurves in soft X-ray (SXR) we use observations from GOES (\citealt{Bornmann1996GOESDisturbances,Chamberlin2009NextSeries,Machol2019GOES-RIrradiance}). GOES includes several geosynchronous spacecrafts flown by the National Oceanic and Atmospheric Administration (NOAA) since 1974. Their main application has been to observe the terrestrial weather and monitor many aspects of the space environment. The solar X-ray Sensor (XRS) measures an incoming solar X-ray emission with a cadence of $1$ s in two channels, $0.5 - 4$ \AA{} and $1 - 8$ \AA{} corresponding to an energy range of $\approx 1.5$ to $\approx 25$ keV.   

\subsection{Methodology}
In this section we first describe the selection criteria of the $73$ flares in this study. We then describe how we use AIA full disk $1600$ \AA{}, and IRIS SJI $1330$ \AA{} and $1400$ \AA{} images to create masks of the flaring ribbon area, and calculate the reconnection flux and its rate for each flare in our sample including their uncertainties. Finally, we present the Wavelet Transform analysis \citep{Torrence1998,Auchere2016ONSERIES} used to characterize and identify the most significant oscillating modes/frequencies for different time series in our data.
 
\subsubsection{Data Selection}
We select $73$ flares from the \verb+RibbonDB+ database \citep{Kazachenko2017} of various flare classes: 10 X-class, 24 M-class, and 39 C-class flares observed from 2010 - 2016. These events are selected manually and represent a sub-sample from more than $3000$ flares in the \verb+RibbonDB+ database. We only select events that exhibit oscillations in magnetic reconnection rates, and have co-temporal X-ray emission time series observations from both GOES XRS and Fermi GBM instruments. Specifically, we compare three time series: magnetic reconnection flux/rate proxies, SXR emission and its rate ($0.5 - 4$ \AA{} and $1 - 8$ \AA{}), and HXR emission ($8-15$ keV, $15-25$ keV, $25-50$ keV, $50-100$ keV, and $100-300$ keV).

In addition, for eight events for which there were co-temporal and co-spatial IRIS and SDO observations, we use IRIS SJI observations to examine fine structure in flare ribbon fronts. While these observations have $1.82$ to $3.75$ times higher spatial resolution than the AIA observations, they have limited FOV, often missing the full extension of the flare ribbons observed by AIA. We therefore only use IRIS data to identify the fine structure development for $8$ out of $73$ events included in this study, and not the reconnection flux analysis.

\subsubsection{Estimating Magnetic Reconnection Flux and Reconnection Flux Rates}
To identify the reconnection flux $\Phi(t)$, we use the vertical component of the HMI vector magnetic field maps, and AIA 1600 \AA{} ribbon maps. To identify the ribbon area A$_{rbn}(\Vec{x},t;c)$, we use the procedure described in \cite{Kazachenko2017}, for AIA and IRIS SJI images. Then the reconnection flux

\begin{align}
    \Phi^{(I_c)}(t_k) & = \int dA_{rbn}(\Vec{x},t_k;c) B_n(\Vec{x},t_k) \approx \sum_{i}^{N_i} \sum_{j}^{N_j} \mathcal{M}^{(I_c)}(x_i,y_j,t_k) B_r(x_i,y_j,t_k) \Delta A. \label{eq:recflx}
\end{align}

The numerical approximation of the magnetic reconnection flux at a time $t_k$ is defined as $\Phi(t_k)$. In this numerical approximation $\mathcal{M}^{(I_c)}(x_i,y_j,t_k)$ refers to the cumulative mask at a time $t_k$ of the sequence of AIA or IRIS SJI images, $x_i$ and $y_j$ refer to the horizontal position of a given pixel, $\Delta  A$ refers to the pixel area size, and $I_c$ to the cutoff intensity criteria to identify the flare ribbon pixels. The cumulative mask -- $\mathcal{M}^{(I_c)}(x_i,y_j,t_k)  = \mathcal{M}^{(I_c)}(x_i,y_j,t_{k-1}) \bigcup  \mathcal{N}^{(I_c)}(x_i,y_j,t_k)$ -- is defined as the set of all the flaring pixels in the instantaneous mask $\mathcal{N}^{(I_c)}(x_i,y_j,t_k) \in I \ge I_c$ up to the \textit{k}-th image in the sequence of observations. The IRIS SJI and HMI observation are co-aligned. Since the IRIS SJIs have higher spatial resolution than the HMI maps, we use \textit{SunPy} to reproject the HMI data to match the dimensions of the IRIS SJI.

We use a cumulative flare ribbon mask to account for the dynamic lifetime of individual flaring pixels. Thus, once a pixel has brightened up and satisfies $I(x_i,y_j,t_k) > I_c(t_k)$, it is included as part of the flare ribbon mask, until the end of the event. The cutoff intensity, which determines if a bright pixel belongs to the instantaneous flare ribbon mask, is defined as $I_c(t_k) = median(I(x_i,y_j,t_k)) \cdot  c$. We use the median as it measures the central value of the intensity maps with a lower bias to the flaring pixels, as compared to the mean intensity. The median intensity of each map is used as a proxy for the intensity of the background chromospheric emission. We use the cutoff parameter $c = \{6,10 \}$, obtained from empirical results both using the $1600$ \AA{} AIA, and $1330$ \AA{} and $1400$ \AA{} IRIS SJI intensity maps. With this range of cutoff intensities we define the positive and negative reconnection fluxes and the errors in these estimates, for positive ($+$) and negative ($-$) polarities, respectively:

\begin{align}
    \Phi^{\pm}(t_k) & = \frac{\Phi^{\pm I_6}(k) + \Phi^{\pm I_{10}}(t_k)}{2} \label{eq:recflxF}.\\
    \delta \Phi^{\pm}(t_k) & = \frac{|\Phi^{\pm I_{10}}(t_k) - \Phi^{\pm I_6}(t_k) |}{2}. \label{eq:rflxn}
\end{align}

The reconnection flux rate for each magnetic polarity is then defined as the time derivative of the reconnection flux $\dot{\Phi}^{\pm}(t_k) = \frac{{\Phi}^{\pm}(t_{k+1}) - {\Phi}^{\pm}(t_{k-1})}{t_{k+1} - t_{k-1}}$. The uncertainty in the reconnection rate for any given time can be estimated using the same formalism as for the reconnection flux:

\begin{align}
    \delta \dot{\Phi}^{\pm}(t_k) = \frac{|\dot{\Phi}^{\pm I_{10}}(t_k) - \dot{\Phi}^{\pm I_6}(t_k) |}{2}.\label{eq:rflxru}
\end{align}

Finally, for a more clear comparison of the magnetic reconnection flux and its rate derived from the flare ribbons with the observed SXR and HXR emission we quantify the unsigned reconnection flux and its rate including uncertainty estimates. First, we define both the unsigned reconnection fluxes and rates as the average between the absolute values of the positive and negative quantities:

\begin{align}
    {\Phi}(t_k) = \frac{\Phi^+(t_{k}) + |\Phi^{-}(t_{k})|}{2}, \label{eq:urecf}\\
    \dot{\Phi}(t_k) = \frac{\dot{\Phi}^{+}(t_{k}) + |\Phi^{-}(t_{k})|}{2}. \label{eq:urecr}
\end{align}
We then define the uncertainty of these measurements as the absolute value of the difference between the two:
\begin{align}
    \delta {\Phi}(t_k) = \frac{|{\Phi}^{+}(t_k) - |{\Phi}^{-}(t_k)| |}{2},\label{eq:urecfu} \\
    \delta \dot{\Phi}(t_k) = \frac{|\dot{\Phi}^{+}(t_k) - |\dot{\Phi}^{-}(t_k)| |}{2}.\label{eq:urecru}
\end{align}
In this approach, the uncertainty of the unsigned fluxes and rates captures the discrepancy between reconnection fluxes and rates at any given time between the opposite magnetic polarities. Ideally, the values of the reconnection flux and their rates should be equal through the full flare evolution since they correspond to the same flaring loop system. The offset represents the uncertainty contributions from the flare ribbon masking algorithm, and the vector magnetic field observations.

\subsubsection{Quantifying Quasi-Periodic Pulsations}
Similar to other astrophysical processes (e.g. \citealt{Aschwanden2011Self-OrganizedAstrophysics}), the Fourier power spectrum of the magnetic reconnection rates follows a power law distribution with frequency, which makes identifying modes of oscillation from the Fourier power spectrum complicated. Further complication arises due to non-stationary oscillations, oscillations with period drifts, which are not accurately described with Fourier Transform methods (e.g. \citealt{Inglis2015,Inglis2016AFLARES}). Thus, to accurately describe a flare oscillation we need a method which can identify the distribution of power among oscillatory modes and how they change in time. The Wavelet Transform accounts for both of these properties and thus is able to identify the most significant stationary and non-stationary oscillations in flare observations, making the Wavelet Transform suitable for studying the statistical properties of flare related oscillations \citep{Broomhall2019}.   

In this paper we use the Wavelet Transform analysis developed by \cite{Torrence1998} and modified by \cite{Auchere2016ONSERIES}. This method has been applied to detect QPPs in solar (\citealt{Clarke2021Quasi-periodicFlare,Kou2022MicrowaveSheet}) and stellar flares (\citealt{Lopez2019}), space weather variability \citep{Richardson2010Near-EarthProperties}, astroseismology \citep{Mathur2010DeterminingStars}, and Earth-climate variability \citep{Webster1998Monsoons:Prediction,Moy2002VariabilityEpoch}. Below we present a brief summary of the Wavelet Transform analysis applied to flare time series $S(t_k)$:
\begin{enumerate}
	\item Shift the mean of the original time series to zero and normalize it by the standard deviation $\hat{S}(t_k) = \frac{S(t_k) - \mu_S}{\sigma_S}$. Here $\hat{S}$ is the normalized time series, $\mu_S$ is the mean of the original time series, and $\sigma_S$ is the variance of the original time series.
	\item Apply the continuous Morlet Wavelet with a nondimensional frequency $\omega_0 = 6$, which is normalized at each frequency such that the energy is unity.
	\item Calculate the regions of the Wavelet Spectrum affected by the finite-length, zero-padded boundaries, called the Cone-of-Influence (COI).
	\item Model the background wavelet transform power and time-averaged wavelet power spectrum as colored noise following a power law distribution inside the COI.
	\item Use this background spectrum to evaluate which signatures of the wavelet transform power and the time-averaged wavelet power spectrum exceed the 95\% significance level of not being attributed to colored noise.
	\item Identify the period (the QPP period) in which the time-averaged wavelet power spectrum exceeding the 95\% colored noise significance level has a local maxima.  
	\item Calculate the Fourier Power spectrum of the normalized time series, and identify the period with the local maximum of Fourier power exceeding the 95\% significance level of colored noise.
	\item Repeat all of the above steps for the fluctuating component of the time series $\delta S(t_k) = S(t_k) - \bar{S}(t_k)$, where $\bar{S}(t_k)$ is the slow varying or background part of the time series. The background time series is calculated using a Savitzky–Golay filter with a 5 minute window.
	\item Compare the QPP periods obtained from the original ($S$) and fluctuation ($\delta S$) time series.
\end{enumerate}

%%%%%%%%%%%%%%%%%%%%%%%%%%%%%%%%%%%%%%%%%%%%%%   	 
\section{Results}\label{sec:res}
In this section we first show two illustrative examples of the spatiotemporal evolution of flare ribbons as observed with AIA 1600 \AA{}, IRIS SJI $1330$ \AA{} and $1400$ \AA{}. We then present the evolution of reconnection fluxes and their rates calculated using the HMI photospheric vector magnetograms and compare these with X-Ray emission measured by Fermi and GOES. Specifically we compare the oscillation periods for each observation, and the delay between the reconnection rate and X-ray emission oscillations. Finally, we present the results of our statistical analysis for $73$ flares with co-temporal SDO, Fermi GBM, and GOES XRS observations.

\subsection{Flare Case Studies with Coordinated SDO, IRIS, Fermi, and GOES Observations} \label{sec:ex}

\subsubsection{Reconnection Rate Oscillations: X1.6 Flare in AR 12158}\label{sec:rec_rate_osc}

\begin{figure}[!t]
	\centering
	\includegraphics[width = \textwidth,trim={4.6cm 5.3cm 4cm 2.9cm}, clip]{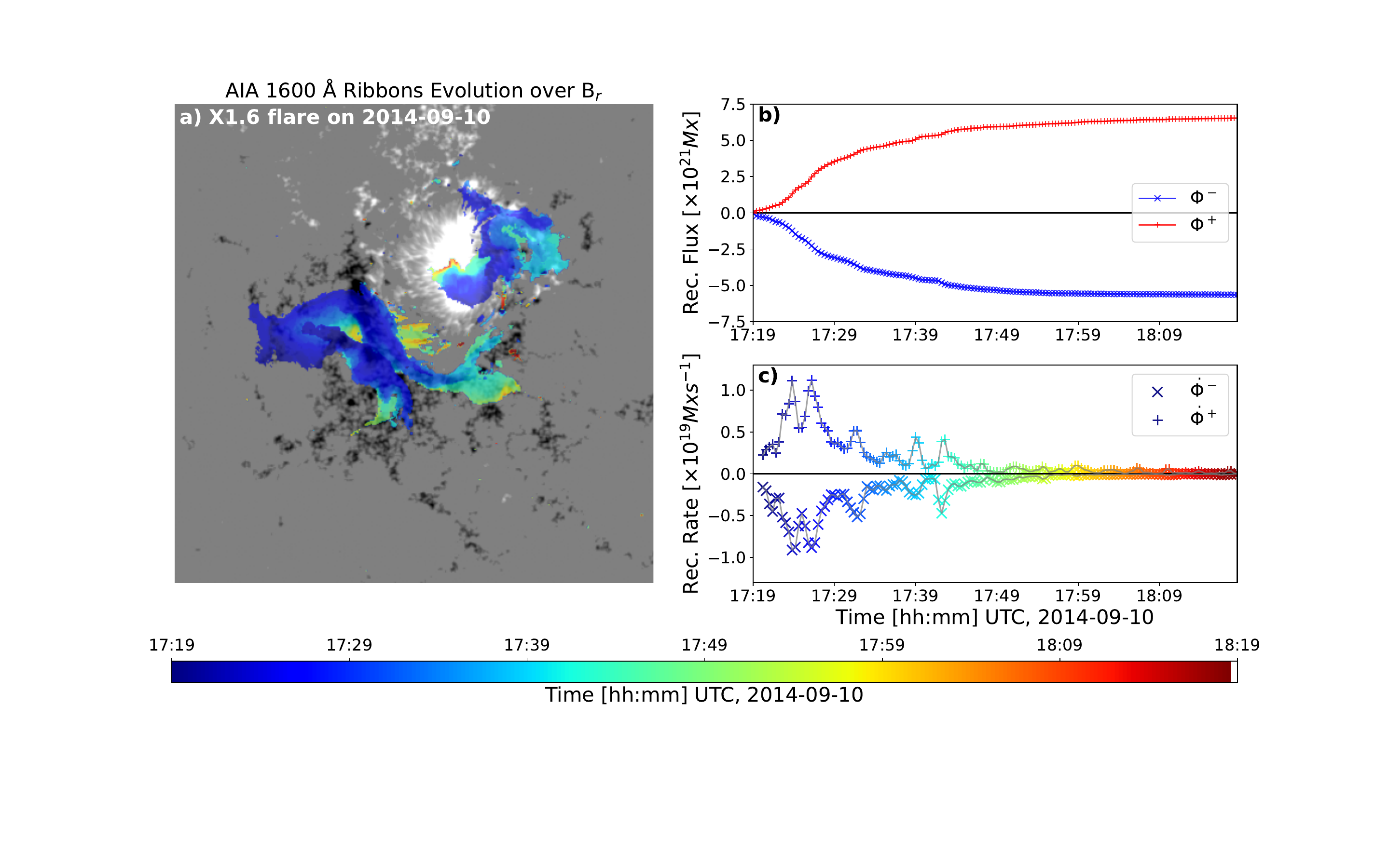}
	\caption{Evolution of flare ribbons and reconnection flux/rate during an X1.6 flare on 2014-09-10 using HMI and AIA $1600$ \AA{} observations. \textbf{a)} Cumulative flare-ribbon mask, colored from early evolution (blue) to the end of the flare (red), over the vertical magnetic field B$_r$ in gray-scale and saturated at $\pm 1000$ G. \textbf{b)} Reconnection flux proxy within positive (red +) and negative (blue -) polarity flare ribbons. \textbf{c)} Reconnection rates for each of the flare ribbons corresponding to each magnetic polarity, the color corresponds to the flare ribbon cumulative mask evolution. See \S~\ref{sec:rec_rate_osc} for more details.}
	\label{fig:aia_ribbon_rec_rate}
\end{figure}

In the left panel of Figure \ref{fig:aia_ribbon_rec_rate} we show the spatiotemporal evolution of flare ribbons in an X1.6-class flare in AR 12158 observed for one hour on 2014-09-10 from 17:19 to 18:19 UTC. The co-temporal full-disk observation obtained from SDO allowed us to track the evolution of the flare ribbons as they swept the active region NOAA 12158. The vertical photospheric magnetic field shown in gray-scale is saturated at $\pm 1000$ G for each corresponding polarity. The top right panel of Figure \ref{fig:aia_ribbon_rec_rate} shows how the cumulative reconnection fluxes in positive and negative polarities increased to $6.54 \times 10^{21}$ Mx and $-5.65 \times 10^{21}$ Mx, respectively. The bottom right panel in Figure \ref{fig:aia_ribbon_rec_rate} shows the evolution of the reconnection rate. The reconnection rate are maximum in each polarity correspond to $11.17 \times 10^{18}$ Mx~s$^{-1}$ and $-9.30 \times 10^{18}$ Mx~s$^{-1}$. As ribbons evolved, the reconnection rate exhibited a series of oscillations of similar magnitudes in both polarities. These bursts gradually decayed as the flare reached its end. We found that larger peaks in the reconnection rates occurred earlier in  the flare at $t < 17$:34~UTC (blue) when the flare ribbons sweep the majority ($>70\%$) of the reconnection flux. Finally, we estimated the unsigned reconnection flux within each burst as $\Phi_{B}(l) = \Phi(t_{l,\mathrm{end}}) - \Phi(t_{l,\mathrm{onset}})$, where $t_{l,\mathrm{onset}}$ and $t_{l,\mathrm{end}}$ are the onset and end times of each reconnection rate burst $l$. We found that for this flare the distribution of reconnected fluxes in each burst ranges from $0.2 \times 10^{20}$ Mx to  $6\times 10^{20}$ Mx.

\subsubsection{Fine Structure of Flare Ribbons: X1.6 and M3.7 Flares}\label{sec:iris_ind}
\begin{figure}[!t]
	\centering
	\includegraphics[width = \textwidth,trim={0cm 4.5cm 0cm 4.5cm}, clip]{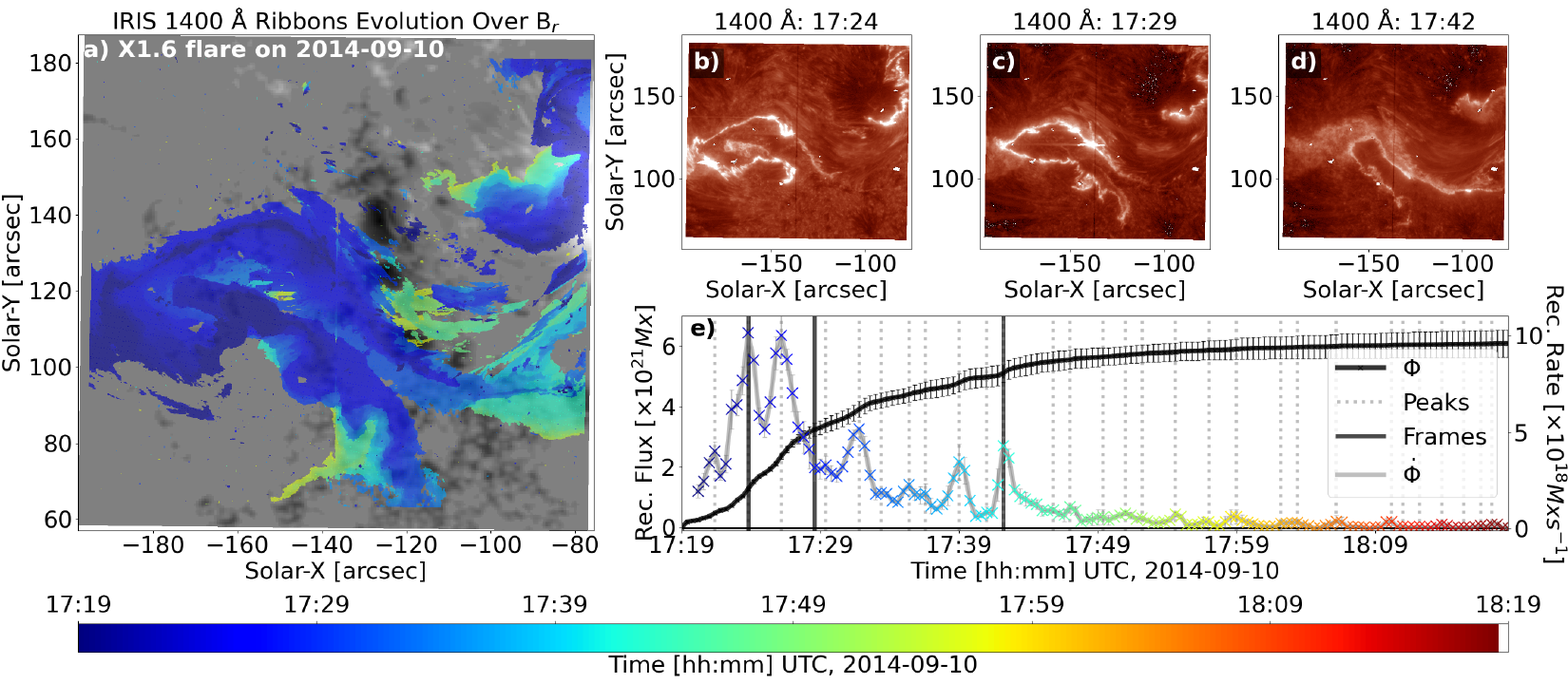}
	\caption{Evolution of flare-ribbon fine structure during an X1.6 flare in AR 12158 (see also Figure \ref{fig:aia_ribbon_rec_rate}) using IRIS $1400$ \AA{} SJI observations. \textbf{a)} IRIS $1400$ \AA{} cumulative flare-ribbon mask evolution over HMI B$_r$. \textbf{b)} IRIS SJI observations of flare ribbons during the reconnection rate maximum at $t \approx$~17:24~UTC. \textbf{c)} Flare ribbons during the decay of the second largest reconnection rate burst at $t \approx$~17:29~UTC. \textbf{d)} Flare ribbons during a later reconnection rate burst at $t \approx$~17:32~UTC. \textbf{e)} Unsigned reconnection flux (black) and its rate (gray). The dotted-vertical gray and solid-vertical black lines represent the reconnection rate peaks and the IRIS observations frames shown in panels (b)-(d) respectively. See \S~\ref{sec:iris_ind} for more details.}
	\label{fig:20140910_iris_rr}
\end{figure}

\begin{figure}[!t]
	\centering
	\includegraphics[width = \textwidth,trim={0cm 4.5cm 0cm 4.5cm}, clip]{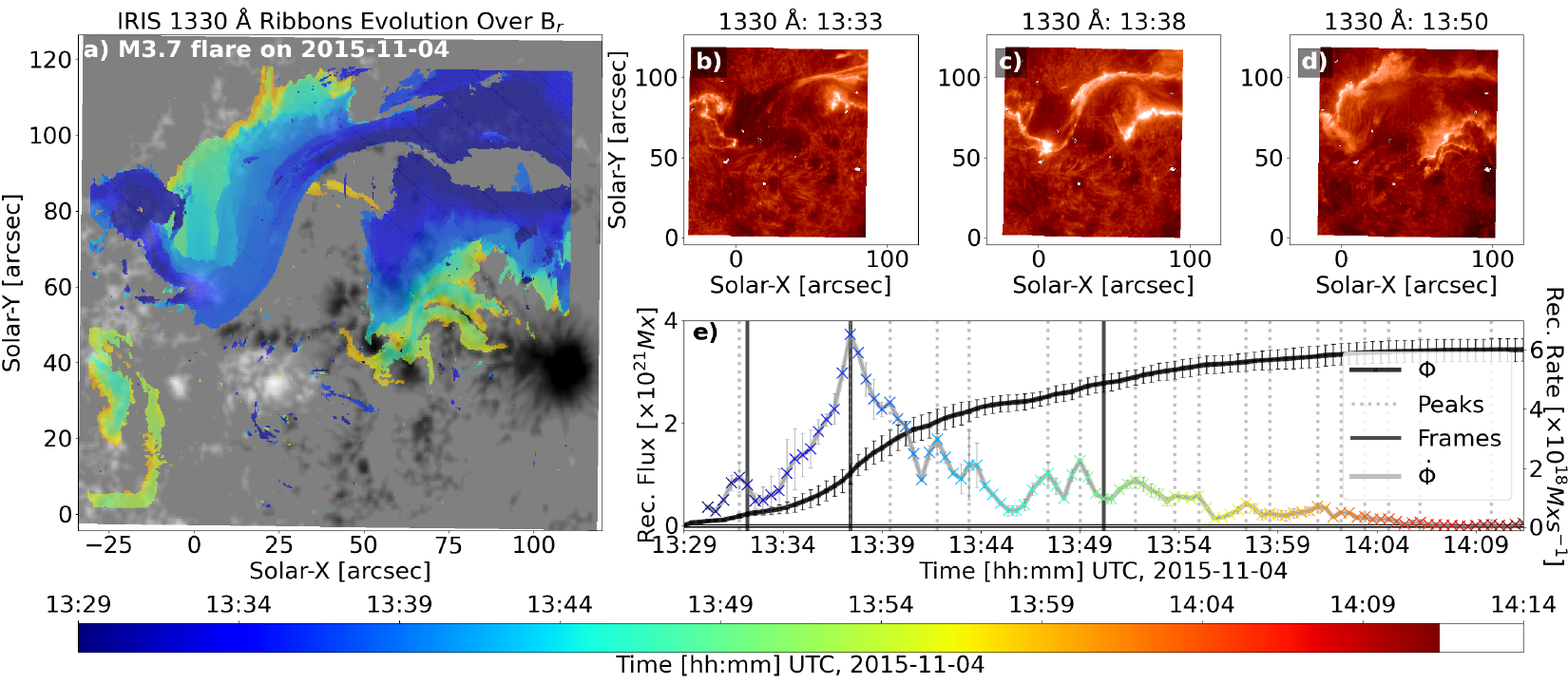}
	\caption{Evolution of flare-ribbons fine structure  during an M3.7 flare on 2015-11-04 in AR 12443 using IRIS $1330$ \AA{} SJI observations. \textbf{a)} Cumulative flare-ribbon mask evolution over B$_r$ as in Figure \ref{fig:20140910_iris_rr}. \textbf{b)} Flare ribbons before the reconnection rate maximum  $t \approx 01:33$ UTC. \textbf{c)} Flare ribbons at the reconnection rate maximum $t \approx 01:38$ UTC. \textbf{d)} Flare ribbons during the decay of a reconnection rate burst at $t \approx 01:42$ UTC. \textbf{e)} Unsigned reconnection flux and rate. See \S~\ref{sec:iris_ind} for more details.}
	\label{fig:20151104_iris_rr}
\end{figure}

Figure~\ref{fig:aia_ribbon_rec_rate} illustrates how  evolution of the reconnection rate is strongly related with the motion of flare ribbons and ribbons spatial structure.
To understand the details of the relationship between the flare ribbon spatial structure affects the reconnection rates we use high-resolution and high-cadence IRIS SJI observations. Specifically, we focus on understanding the connection between the flare-ribbon fine structure, which we use to refer to the complexity in the form of waves and swirls on the ribbon fronts, and the bursts in the reconnection rates. Figures \ref{fig:20140910_iris_rr}(a) and \ref{fig:20151104_iris_rr}(a) show evolution of the cumulative flare-ribbon mask derived from the IRIS SJI observations for the same X1.6 flare on 2014-09-10 as shown in  Figure \ref{fig:aia_ribbon_rec_rate}, and the M3.7 flare observed on 2015-11-04, respectively. The 2014-09-10 flare is observed in the $1400$ \AA{} filter with a spatial resolution of  $0.166' \times 0.166'$ and a 19 second cadence, while the 2015-11-04 flare is observed in the $1330$ \AA{} filter with a $0.332' \times 0.332'$  resolution and a 12 second cadence. These high-resolution flare-ribbon masks are compared to the average unsigned reconnection flux and its rate calculated using AIA $1600$ \AA{} observations.   

First, we analyze the X1.6 flare described in \S \ref{sec:rec_rate_osc}. Figure \ref{fig:20140910_iris_rr} has a smaller FOV than the cutout from the full solar disk observations of SDO. This limits using the SJI observations to calculate the magnetic reconnection flux/rate, since they rarely capture the full spatial evolution of flare ribbons. We therefore continue using AIA $1600$ \AA{} data to calculate our reconnection flux/rate proxies. Yet, the high spatial resolution SJI observations from IRIS show fine structure details within the flare ribbon fronts. The SJIs in Figures \ref{fig:20140910_iris_rr}(b-d) highlight how the flare ribbon evolves in response to bursts in the reconnection rates. The SJIs display fine structure in the form of swirls and waves which emerge and decay through the lifetime of the reconnection rate bursts. These fine-structure patterns contribute to the large-scale bursty evolution of the flare ribbon, as ribbons sweep new magnetic flux from the active region. For the first time, we relate bursty sweeps in the spatial evolution of flare ribbons to bursts in the magnetic reconnection rate, which temporally evolve as quasi-periodic oscillations.

The co-temporal development of fine structure and quasi-periodic reconnection rate burst is also observed in Figure \ref{fig:20151104_iris_rr} for the M3.7 flare on 2015-11-04. The ribbon masks derived from IRIS observations (Figure \ref{fig:20151104_iris_rr}(a)) show that during the early stages of the flare ($t < 13$:45~UTC), the flare ribbon has swept more than 50\% of the total reconnection flux in 5 reconnection rate bursts. Similar to Figure \ref{fig:20140910_iris_rr}(b-d), the $1330$ \AA{} IRIS SJIs show the co-development of fine structure and reconnection-rate bursts. We also find that fluxes swept by individual bursts range from $0.5\times 10^{20}$ Mx to $6.5\times 10^{20}$ Mx. The co-temporal development of flare-ribbon fine structure, observed in $1330$ \AA{} and $1400$ \AA{} passbands, and the reconnection rate QPPs suggests that these phenomena could be related to the dynamics of magnetic reconnection and conditions of the current sheet.

We propose that the fine-scale structures that are observed in the high-spatial resolution IRIS SJI observations are related to processes and structures higher in the corona where magnetic reconnection with the flare loops is happening (i.e. the current sheet; \citealt{Brannon2015SPECTROSCOPICWAVES,Wyper2021IsSheet}). In section \ref{sec:dis} we compare this interpretation and others with our findings of magnetic-reconnection rate bursts in more detail.

\subsubsection{Comparison of Fluctuations in the Reconnection Rate with Fluctuations in the X-ray flux: M3.7 and M2.1 Flares}\label{sec:rec_xray}
Until now we have focused our attention on the spatial complexities of flare ribbons and their relationship with QPP bursts in the magnetic reconnection rates. In this section we compare temporal evolution of the reconnection flux rate to the emission of HXR light curves. Specifically we study two flares, the M3.7 flare on 2015-11-04 and M2.1 flare on 2012-03-06.

\begin{figure}
\gridline{\fig{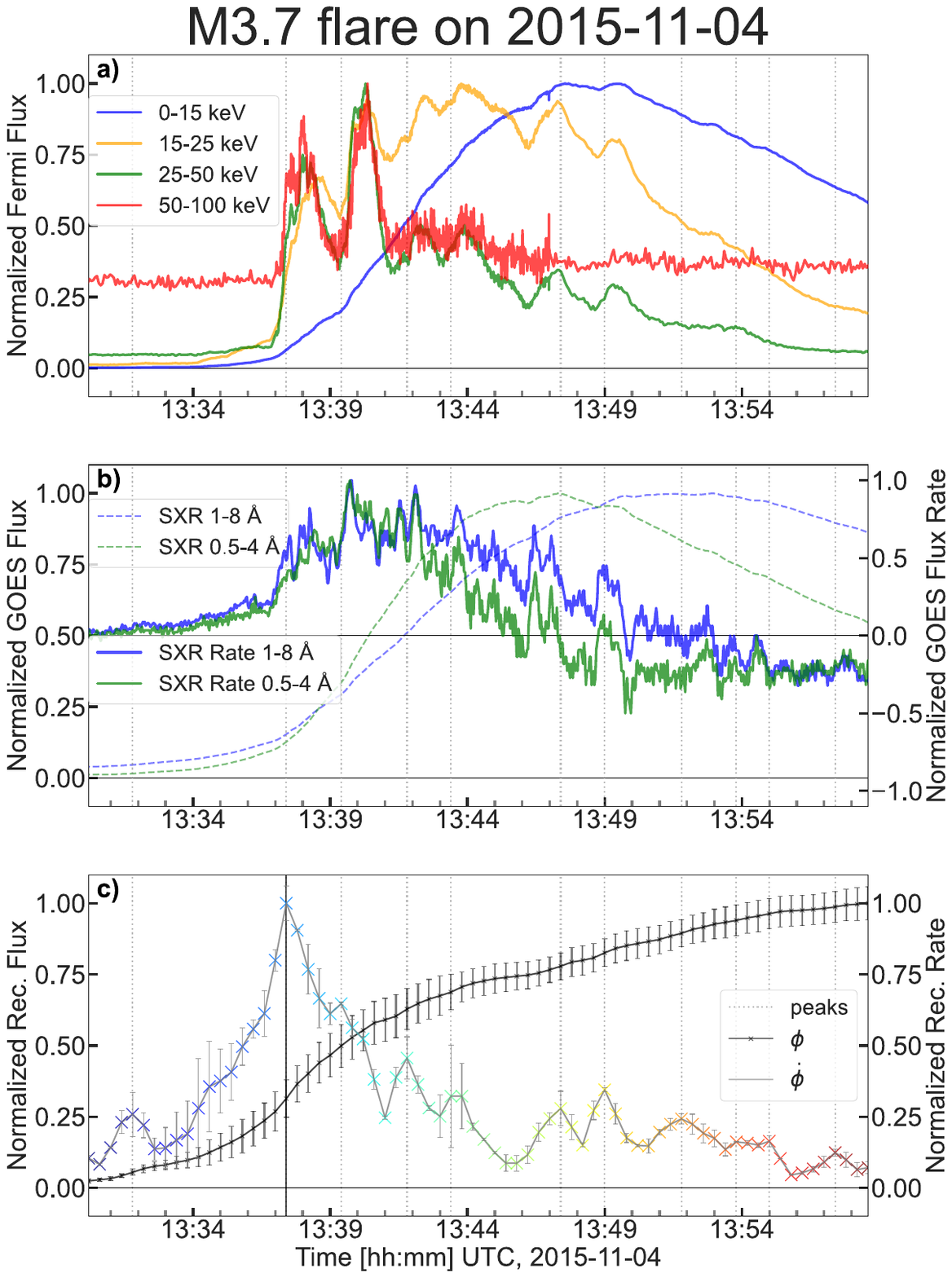}{0.49\textwidth}{}
\fig{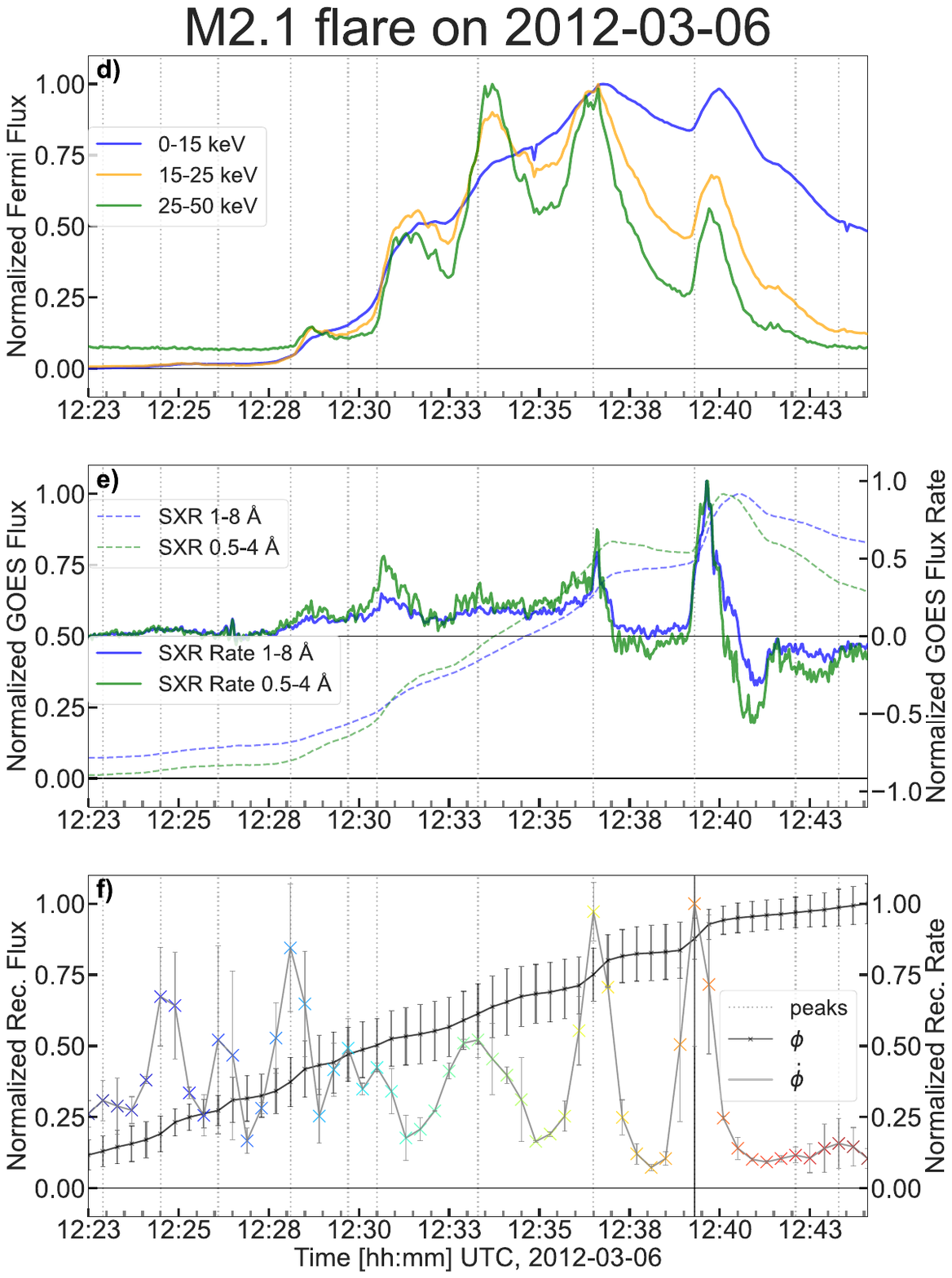}{0.49\textwidth}{}}
\caption{Comparison of fluctuations in magnetic reconnection rate with X-ray fluxes from Fermi GBM and GOES XRS during two M-class flares. The \textit{Left} panel shows the M3.7 flare (see also Figure \ref{fig:20151104_iris_rr}). The \textit{Right} panel shows the M2.1 flare observed on 2012-03-06. \textbf{a)} Fermi GBM flux observed in four energy bands: $8-15$ kev (blue), $15-25$ keV (orange), $25-50$ keV (green), and $50-100$ keV  (red). \textbf{b)} Solid black and dotted gray lines show the GOES SXR fluxes; blue and green lines show  and GOES SXR fluxes time derivatives for $1-8$ \AA{} and $0.5-4$ \AA{} bands, respectively. \textbf{c)} Reconnection flux (black) and its rate (colored to scale with temporal evolution). All of the variables have been normalized by their respective maximum. The vertical black lines mark peaks in the magnetic reconnection rate. See \S~\ref{sec:rec_xray} for more details.}
\label{fig:xray_rec_ind}
\end{figure}

In Figure \ref{fig:xray_rec_ind} we show examples of X-ray light curves from Fermi GBM and GOES XRS for two flares (out of $73$): an M3.7 flare on 2015-11-04 (also shown in Figure \ref{fig:20151104_iris_rr}) and an M2.1 flare on 2012-03-06. We only include those Fermi channels that are not affected by noise, and where observations from two GOES channels are available. We further use the derivative of the GOES SXR emission as a proxy for the HXR emission via the Neupert effect \citep{Neupert1968ComparisonFlares}.  

Both flares in Figure \ref{fig:xray_rec_ind} serve as an illustrative example of the delay that we find between the temporal profiles of the X-ray emission and its rate and the magnetic reconnection rate. The cross-correlation between the fluctuating components of the magnetic reconnection rates and the X-ray emission yields delays within $2$ minutes. Here the positive delays mean that the reconnection rate QPP happens before the X-ray QPPs, and the negative is vice-versa. Interestingly, the lower amplitude bursts in the reconnection rates are not always accompanied by a corresponding burst in the HXR emission. This is especially noticeable comparing early phases of the Fermi GBM with the reconnection rate profiles for the 2012-03-06 flare, shown in Figure \ref{fig:xray_rec_ind}(d-f). For this flare the first Fermi GBM X-ray bursts are observed only 7 minutes after the flare onset, and more than five minutes after the initial reconnection rate burst ($t \approx 12$:24:00~UTC). Additionally, during the late phase ($t>13$:50:00~UTC) of the 2015-11-04 flare, shown in Figure \ref{fig:xray_rec_ind}(a-c), the high energy Fermi channels ($>25$ keV) do not exhibit the same busts as those in the reconnection rate. In Figure~\ref{fig:rdb_exp} of Appendix~\ref{sec:sup_mat} we show  additional 10 examples comparing Fermi HXR and GOES SXR QPPs with the reconnection rate oscillations. These additional 10 examples illustrate that QPPs observed in both X-ray and the reconnection rate are not unique to the two case studies shown here potentially being a consequence of the same process in the current sheet.

\begin{figure}[!ht]
 	\centering
 	\includegraphics[width=0.9\textwidth,trim={0cm 1.7cm 0cm 1.7cm}, clip]{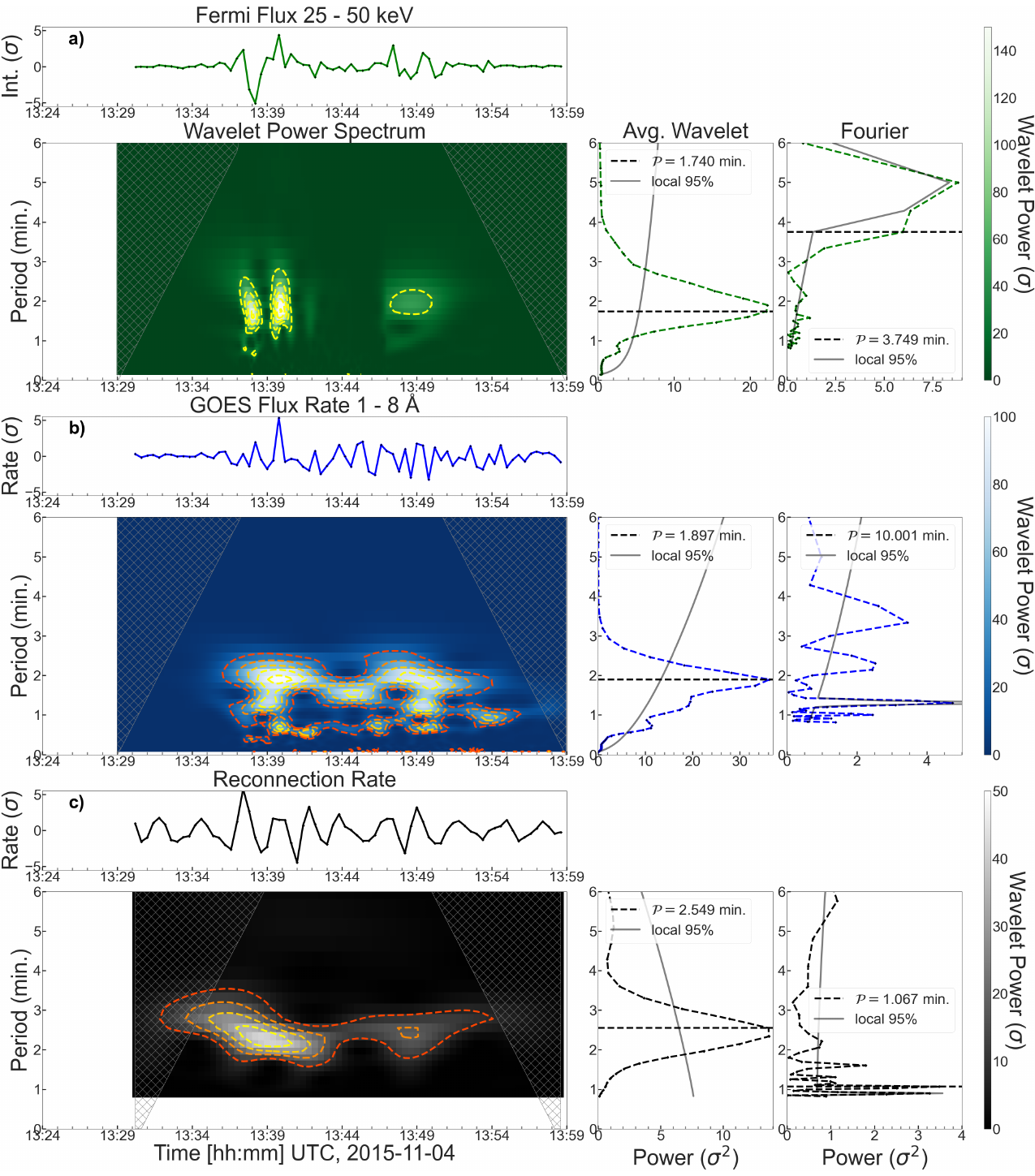}
 	\caption{Wavelet and Fourier power spectra for X-ray fluxes and reconnection rate for the M3.7 2015-11-04 flare (see also Figures~\ref{fig:20151104_iris_rr} and \ref{fig:xray_rec_ind}(a-c)). \textbf{Panel a)} The detrended fluctuations in the Fermi $25-50$ keV emission are shown on the top panel, its wavelet transform below it, and average wavelet power and Fourier power on the right. \textbf{Panel b)} Fluctuation in the GOES $1 - 8$ \AA{} emission rate following the same layout as in (a). \textbf{Panel c)} Fluctuations in the magnetic reconnection rate  with the same layout as in (a) and (b). The red to yellow contours on the wavelet transform maps correspond to areas where wavelet power exceeds the $95$\% colored noise level. See \S~\ref{sec:rec_xray} for more details.}
	\label{fig:wavelet_example}
\end{figure}

Lastly, in Figure \ref{fig:wavelet_example}  we apply the wavelet transform to characterize the QPPs in the 2015-11-04 flare using emission in the $25 - 50$ keV,  emission rate in the $1-8$ \AA{}, and the magnetic reconnection rate. The background-removed HXR and the SXR rate QPPs have very similar periods of approximately 1.8 minutes, while the reconnection rate has a higher period of 2.5 minutes. The periods in the reconnection rate and X-Ray QPPs are not the same, which is expected due to the discrepancies in the time-series. Nonetheless, the wavelet power spectrum for these observations shows that similar oscillatory modes are excited co-temporally in all of the time-series. We compare these results with the Fourier power spectrum and find inconsistency in the periods with the local maxima between the different observations, and the wavelet results. These periods also do not match the observed oscillation periods. The Fourier power spectrum is excluded from the rest of the study since it fails to accurately represent the power of non-stationary oscillations. Furthermore, the Fourier power spectrum distribution of local maxima makes it much harder to identify QPPs that match oscillations identified in the time series data.

The co-temporal bursty signatures in both of the HXR emission observed by Fermi GBM, GOES XRS SXR emission and the reconnection rates from SDO suggest a link between the two types of observations. This link is further demonstrated by the results of the wavelet transform: the co-temporal excitation of similar oscillatory modes ($\mathcal{P}\le 3$ minutes), and finding QPPs in the time-averaged wavelet spectrum with similar periods (1.8 and 2.5 minutes). In section \ref{sec:dis} we discuss this relationship in more detail.

\subsection{Statistical Properties of QPPs in Reconnection Rates and HXR Emission for 73 Flares}\label{sec:summary}

To further understand and characterize the relationships that we found for the example events in \S~\ref{sec:ex} we expand our analysis to $73$ flares: $39$ C-class, $24$ M-class, and $10$ X-class flares from the \verb+RibbonDB+ database.

\subsubsection{Comparison of QPP Periods: Reconnection Rates and X-ray Emission}\label{sec:sum-periods}

\begin{figure}[!ht]
	\centering
	\includegraphics[width = \textwidth,trim={0cm 0cm 0cm 0cm}, clip]{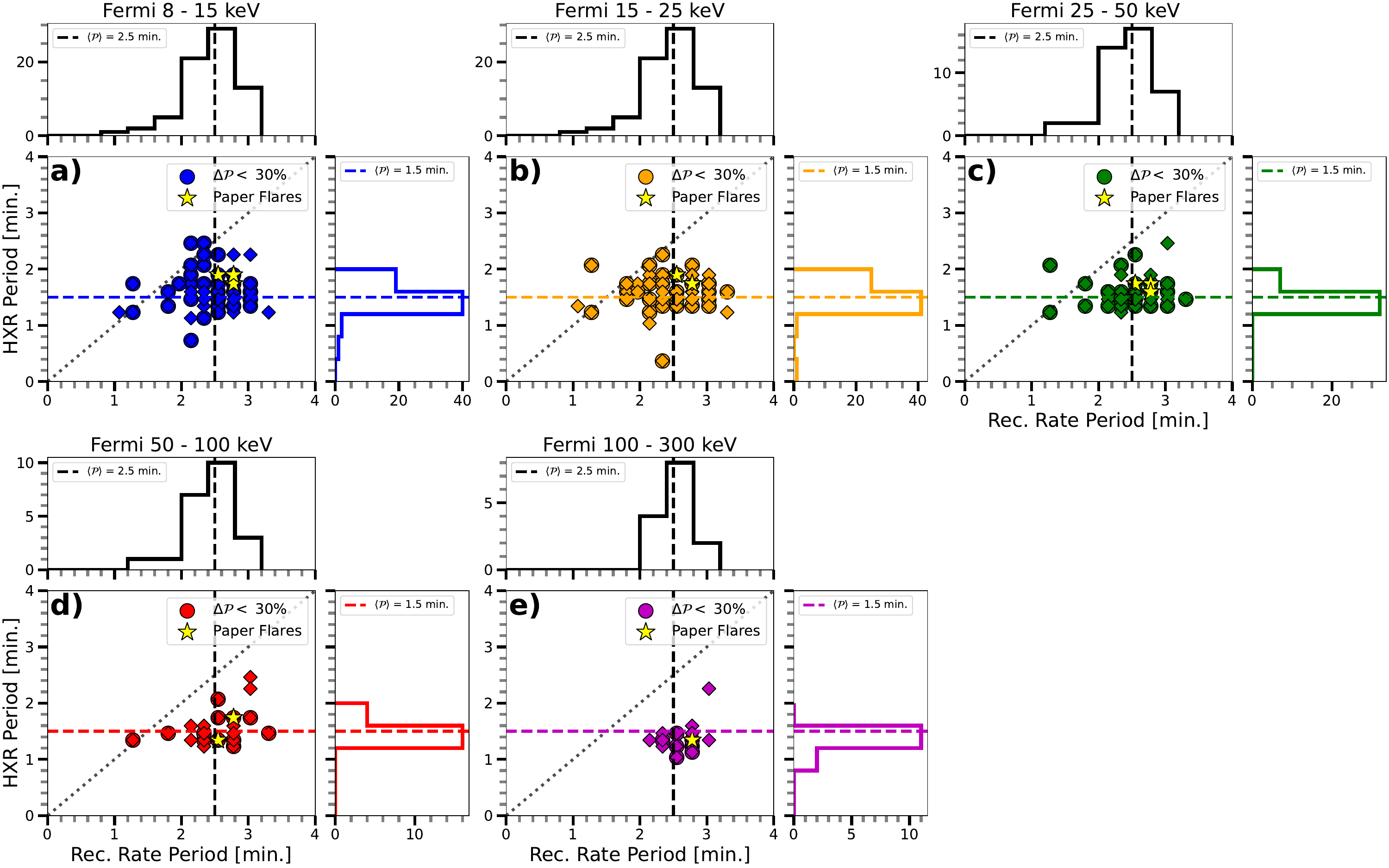}
	\caption{Periods of quasi-periodic pulsations (QPPs)  in HXR emission (from Fermi GBM) vs. periods of QPPs in unsigned magnetic reconnection rates for $73$ flares. Panels a) to e) correspond to the following X-ray energy bands: \textbf{a)} $8-15$ keV, \textbf{b)} $15-25$ keV, \textbf{c)} $25-50$ keV, \textbf{d) }$50-100$ keV; and \textbf{e)} $100-300$ keV. QPPs period have been identified using Wavelet transform. The large circles represent the flares with periods that match up to $30$\% between the original time-series, and the background subtracted time-series. The yellow stars show case studies from \S~\ref{sec:ex}. See \S~\ref{sec:sum-periods} for more details.}
	\label{fig:fermi_rdb_periods}
\end{figure}

We apply the wavelet transform to the unsigned reconnection rates of all $73$ flares to find their periods. We then compare the QPP periods from the unsigned reconnection rates with those derived from the Fermi GBM HXR channels (Figure~\ref{fig:fermi_rdb_periods}). First, the number of QPP observations in each energy channel varies---lower-energy channels ($8 - 50$ keV) have more events than higher-energy channels (panels (a) and (e)). Second, there is no simple mathematical relationship between the observed QPPs in the reconnection rate and the HXR Fermi observations. Rather, the periods converge in the range of $0-3$ minutes. The histograms of the QPPs show that the HXR distribution peaks at around 1.5 minutes in each channel, while the reconnection rate distribution peaks at 2.5 minutes. Therefore, on average the QPPs in the reconnection rate have larger periods than those in the HXR emission. This is the same behavior observed in the example flares (yellow stars in Figure~\ref{fig:fermi_rdb_periods}) shown in \S \ref{sec:ex}.

\begin{figure}[!ht]
	\centering
	\includegraphics[width = 0.7\textwidth,trim={0cm 0cm 0cm 0cm}, clip]{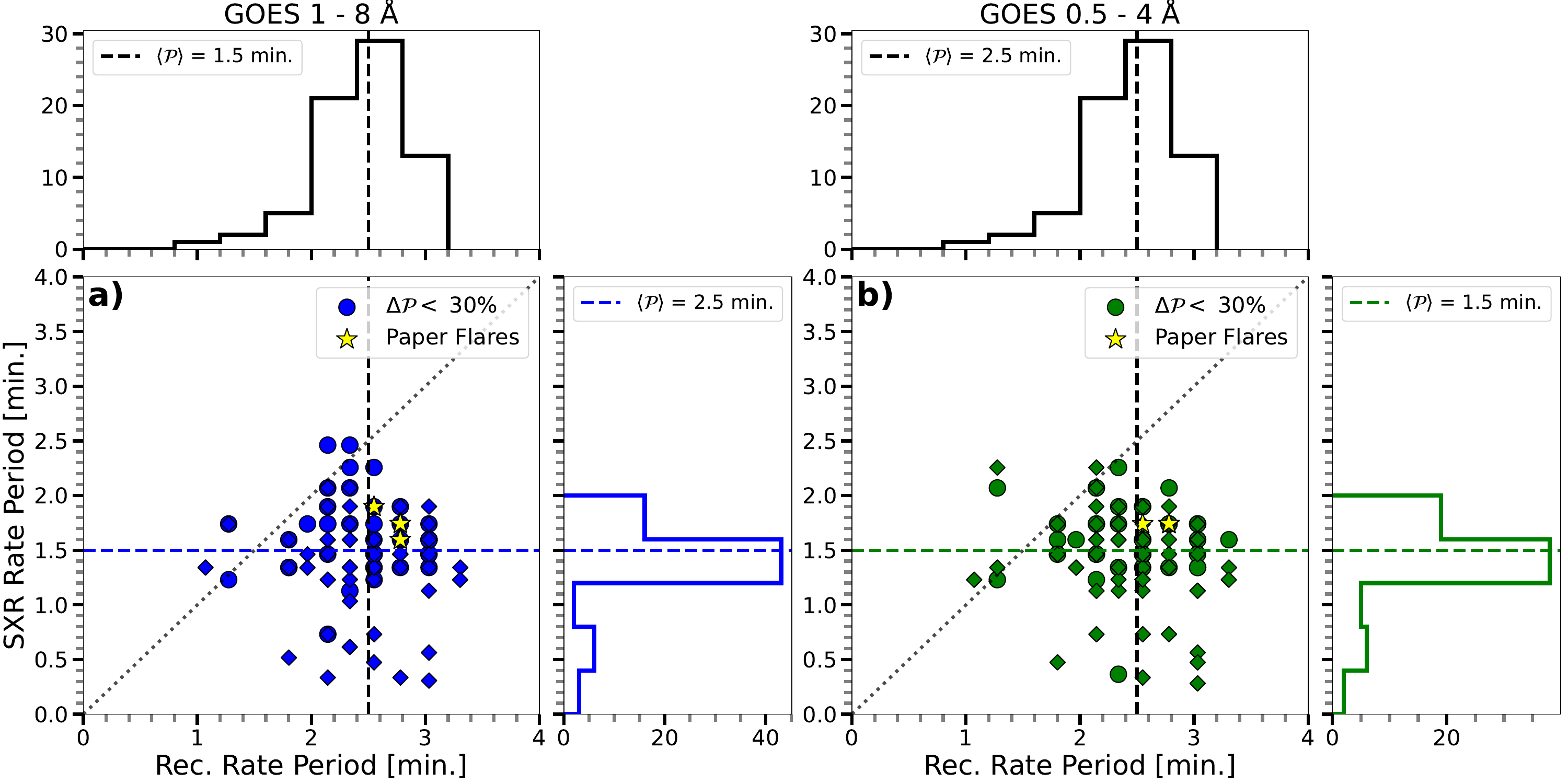}
	\caption{Periods of QPPs in SXR emission rates vs. periods of QPPs in unsigned magnetic reconnection rates for $73$ flares: \textbf{a)} $1-8$ \AA{} and \textbf{b)} $0.5-4$ \AA{} passbands. The large circles represent the flares with periods that match up to $30$\% between the original time-series, and the background subtracted time-series. The yellow stars show case studies from \S~\ref{sec:ex}. See \S~\ref{sec:sum-periods} for more details.}
	\label{fig:goes_rdb_periods}
\end{figure}

In Figure~\ref{fig:goes_rdb_periods} we compare the QPP periods from the unsigned reconnection rate and GOES XRS flux derivative. We find the same number of events in both the GOES long and short wavelength bands, since they are much closer in energy range than the Fermi GBM channels. Similar to Figure \ref{fig:fermi_rdb_periods}, the QPPs in GOES XRS flux-derivative have no polynomial relationship with QPSs in  reconnection rate. Instead we find the same convergence in periods below 3 minutes. Coincidentally, the SXR rate QPPs also have a peak in their distribution at approximately 1.5 minutes. Therefore, as was the case in the Fermi HXR observation, the SXR emission rate QPPs have a shorter period on average than the reconnection rates.
 
Our results suggest that the wavelet transform is capable of characterizing the QPPs in the X-ray and reconnection rate data. This is confirmed by the overall agreement between QPPs observed in the isolated fluctuating component of each time series, and the time series containing background trends. On average the HXR emission and SXR emission rates have the same QPP periods of 1.5 minutes, which is lower than the 2.5 minute QPP periods in the reconnection rates. This statistical relationship suggests that the HXR and SXR QPPs are intrinsically linked to the same driving mechanics related to the reconnection rate QPPs.

\subsubsection{Delay in X-ray Emission}\label{sec:xray-delay}
When examining the individual examples presented in Figure \ref{fig:xray_rec_ind}, we found that both the Fermi GBM and GOES XRS rate oscillations are delayed from the reconnection rate bursts. To evaluate the delay time for each flare we examine the cross correlation between the background subtracted reconnection rate and the X-ray data. The location of the cross correlation maxima determines the delay time between the oscillations. We determine delay calculations relative to the reconnection rate, therefore positive values imply that the reconnection rate oscillation precedes the X-ray oscillations, while a negative value is the opposite.

\begin{figure}[!ht]
	\centering
	\includegraphics[width=\textwidth]{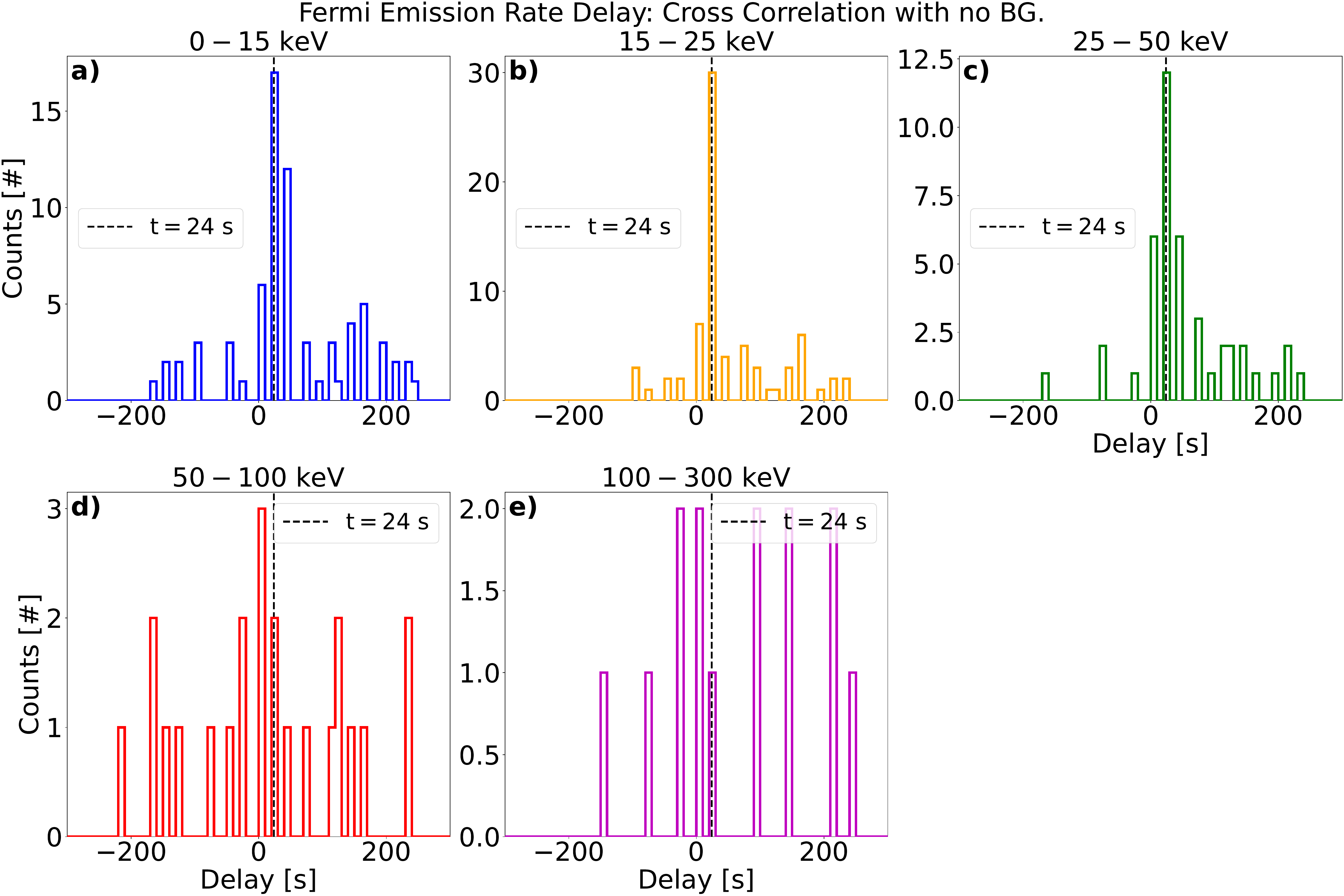}
	\caption{Distributions of delay times between HXR emission and magnetic reconnection rate for $73$ flares. Panels a) to e) correspond to HXR emission  in \textbf{a)} $8-15$ keV, \textbf{b)} $15-25$ keV, \textbf{c)} $25-50$ keV, \textbf{d)} $50-100$ keV and \textbf{e)} $100-300$ keV energy bands. Dashed vertical lines show 24 seconds of delay time from co-temporal emission. Positive and negative delays are defined as reconnection rate QPP preceding and following the X-ray QPP, respectively. See \S~\ref{sec:xray-delay} for more details.}
	\label{fig:fermi_delay}
\end{figure}

In Figure \ref{fig:fermi_delay} we show the delay time of Fermi GBM HXR emission in each channel. Each energy channel shows very similar behavior, highlighting that both the low and high HXR QPPs are connected to the same dynamical process. We see a distribution of delay times ranging from $\pm 300$ seconds. The lower energy channels ($8-50$ keV) show a peak in the distribution of delays at $24$ seconds. This maxima is not present in the higher energy channels ($50 - 300$). Instead we get somewhat of a more uniform distribution of delays. This can be attributed to the low number of events that have clean HXR oscillations in these channels (23 and 14 flares respectively). Therefore, the HXR oscillations sometimes precedes the reconnection rate bursts, but mostly occur after the reconnection rate oscillation by a couple of seconds to minutes. On average, the HXR oscillations occur $24$ seconds after the onset of the reconnection rate QPPs.

\begin{figure}[!ht]
	\centering
	\includegraphics[width=0.75\textwidth]{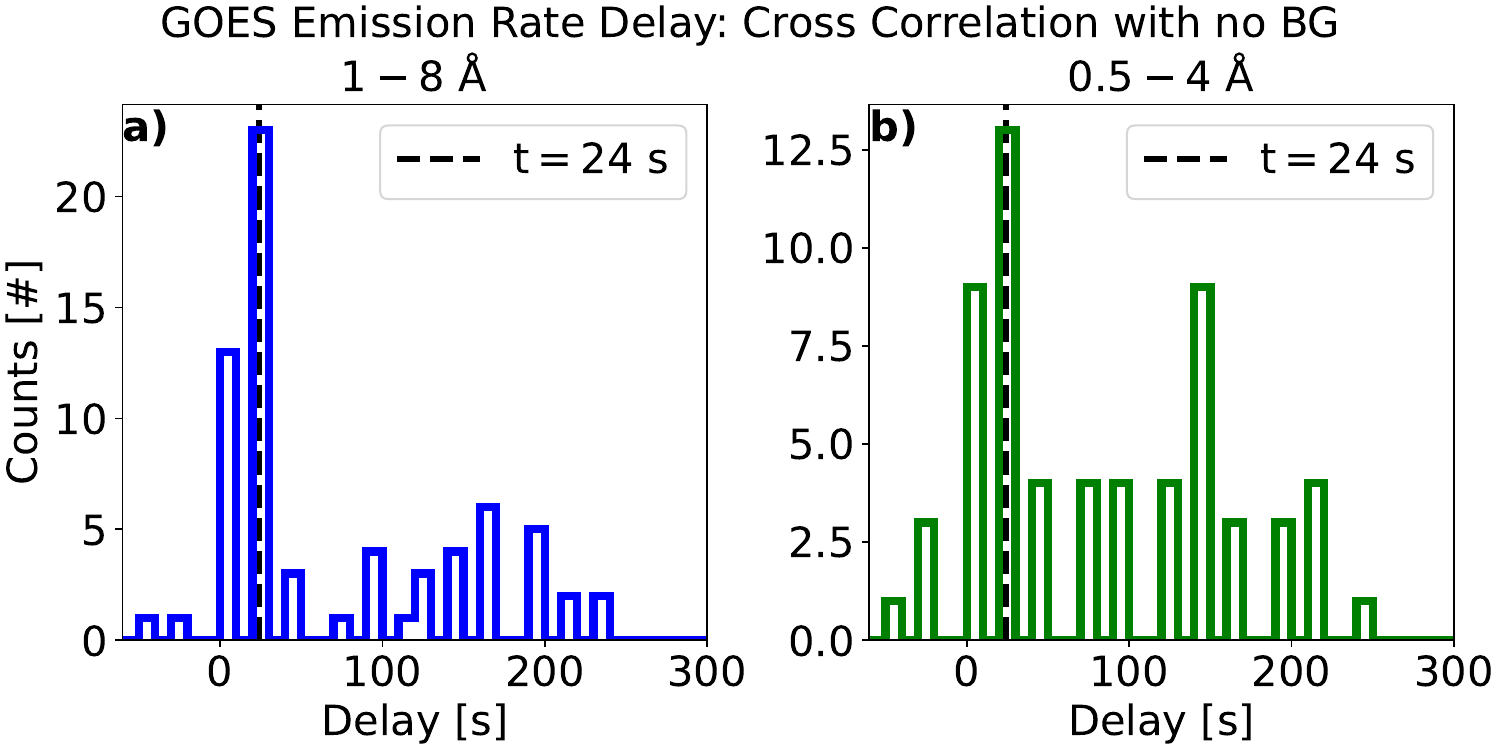}
	\caption{Histograms showing the SXR emission rate  delay from the magnetic reconnection rate for  73 flares. \textit{Left} and \textit{Right} panels show $1-8$ \AA{} and $0.5-4$ \AA{} channels, respectively. See \S~\ref{sec:xray-delay} for more details.}
	\label{fig:goes_delay}
\end{figure}

Figure \ref{fig:goes_delay} presents the cross-correlation time delay time between the GOES XRS emission rate QPPs (HXR emission proxy) and the reconnection rate oscillations. The distributions of delays for both channels ($1-8$ \AA{} and $0.5-4$ \AA{}) are spread from -60 to 300 seconds. Furthermore, these distributions peak at 24 seconds, suggesting, as was the case for the HXR (Figure \ref{fig:fermi_delay}), that the SXR rate oscillations occur after the reconnection rate. Specifically we find that on average, the reconnection rate burst onset precede the SXR rate QPPs by $24$ seconds. Thus the GOES SXR emission rate QPPs have the same average statistical characteristics found for the Fermi HXR QPPs. This is further evidence that although the SXR and HXR emission have different sources (flare footpoints and loop-tops, and flare loops respectively) their QPPs suggest a connection to the same oscillatory particle acceleration episodes.

Since most of the X-ray observations show a peak in their distribution at 24 seconds we further examine if the result is a real phenomenon or an artifact in the data. A delay time of 24 seconds corresponds to the time cadence of the AIA observations used to calculate the reconnection flux and rates. Therefore, the 24 second delay could be an error introduced by the finite difference scheme used to calculate the reconnection rate. We verify if the numerical scheme introduced the delay between the HXR and reconnection rate QPPs by changing the numerical differential scheme from central difference scheme, which is the default of the \textit{NumPy} library \citep{Harris2020ArrayNumPy}, to finite forward and backwards schemes. We find that changes in the numerical approximation of the derivatives cause no significant changes to the location of the peak of the distributions when using forward differentiation. When using the backwards differentiation scheme the peaks in the delay distributions shift to 5 seconds. Therefore, the order of which QPP precedes (reconnection rate) remains unchanged when varying the differentiation scheme. Yet the average value 24 second delay time should be interpreted as an upper limit because of the value's sensitivity to changes in differentiation schemes.	 

Table \ref{tab:sum} summarizes the results of this section. The oscillation periods presented in our statistical analysis range from tens of seconds to four minutes. On average the reconnection rate oscillations have a period of 2.5 minutes, which is larger than the X-Ray average QPP of 1.5 minutes. The reconnection rate QPPs' onset precedes the X-ray QPPs in the lower energy channels ($>50$ keV) on average by 24 seconds. The higher energy X-ray channels ($\ge 50$ keV) QPPs on average occur more synchronously to the reconnection rate QPPs. We note, however, that this result is subject to statistical biases due to the low number of flares (23 and 14 respectively) that exhibit clear QPPs in the higher energy channels.

\begin{deluxetable}{lcccc}
	\tablecaption{Summary of the statistical analysis of QPP periods and delay times in HXR emission for 73 analyzed flares. \label{tab:sum}}
	\tablehead{
	\colhead{Date}&\colhead{Flare start time [UTC]}&\colhead{GOES FLare Class}&\colhead{NOAA AR}}
	\startdata
	Rec. Rates       	& 73 & $2.5$ & N/A \\
	Fermi $8-15$ keV 	& 72 & $1.5$ & 24 \\
	Fermi $15-25$ keV	& 73 & $1.5$ & 24 \\
	Fermi $25-50$ keV	& 43 & $1.5$ & 24 \\
	Fermi $50-100$ keV   & 23 & $1.5$ & 20 \\
	Fermi $100-300$ keV  & 14 & $1.5$ & 16 \\
	GOES $1-8$ \AA{} 	& 73 & $1.5$ & 24 \\
	GOES $0.5-4$ \AA{}   & 73 & $1.5$ & 24
	\enddata
\end{deluxetable}

\section{Discussion}\label{sec:dis}

Our spatiotemporal analysis of flare ribbons and reconnection rate time series showed that magnetic flux swept by flare ribbons produces bursts or oscillations in magnetic reconnection rates. Bursty magnetic reconnection is associated with the plasmoid instability (PI), and the formation and dynamics of plasmoids of different scales within the current sheet \citep{Shibata2001Plasmoid-induced-reconnectionReconnection}. Analytical work \citealt{Wyper2021IsSheet} suggest that these swirls and waves and swirls shown in Figures \ref{fig:20140910_iris_rr} and \ref{fig:20151104_iris_rr} could be related to the PI. In particular, \cite{Wyper2021IsSheet} used an analytical representation of an eruptive flux rope and flaring loop system to show that plasmoid structures and their motions within the current sheet were connected to the fine structure and their apparent motion displayed in the flare ribbons.

\cite{Brannon2015SPECTROSCOPICWAVES} analyzed the M7.3 class flare on 2014-04-18 finding QPP signatures in the position and Doppler shifts of sawtooth fine structure (wave breaks) patterns. These oscillations were found to have periods periods of $140$ seconds and $100$ to $200$ seconds respectively. Coincidentally, these are within the period range of detected QPPs from the reconnection rates in our sample of $73$ flares ($24-240$ seconds). Specifically, we found that the M7.3 on 2014-04-18 flare had a reconnection rate QPP of $140$ seconds, which agrees with the authors findings. \cite{Brannon2015SPECTROSCOPICWAVES,Parker2017ModelingShear} both suggested that a possible explanation of the oscillations found in the motions of the fine structure in the flare ribbon was the PI in the above the loop-top region of the flare system.

Using IRIS SJIs in the $1330$ and $1400$ \AA{} filters we found fine structure in the flare ribbon fronts evolving throughout the flares. We found that the bursty quasi-periodic evolution of the reconnection rate occurred co-temporally with the development and evolution of fine structure in the flare ribbon fronts. This behavior is presented in Figures \ref{fig:20140910_iris_rr} and \ref{fig:20151104_iris_rr}. Yet, for now we did not isolate the contributions of the flare-ribbon fine structure from the cumulative evolution of flare ribbon. This step will be necessary to explore the contribution of the fine structure to the reconnection rate bursts. A future study will be conducted to explore this relationship, and uncover current sheet dynamics that contribute to the formation and evolution of the flare-ribbon fine structure. Furthermore, in this study we used the AIA $1600$ \AA{} observations to derive the reconnection flux and rate. The lower spatial resolution and saturation issues of this passband complicated the identification of the location of the fine structure development in our observations. Thus, to explore the relationship between the fine structure evolution and the reconnection rate burst we will need a data set with very high spatial and temporal resolution. To summarize, we find that the larger-scale cumulative flare ribbon evolution, which contains the regions in which the fine structure develops, results in a bursty profile of the reconnection rate as shown in Figure \ref{fig:aia_ribbon_rec_rate}.

We discard the possibility that bursts in the reconnection rate are a consequence of a systemic bias in our masking algorithm, since we find reconnection rate QPPs in both positive- and negative-polarity ribbons independently. For example, Figure \ref{fig:aia_ribbon_rec_rate} shows that although the morphology of the flare ribbon is very different, the reconnection flux/rate approximately balance each other in the opposite polarities through the flare. Additionally, Figure \ref{fig:rdb_exp} shows $10$ examples of the reconnection flux and rates for the positive and negative polarity ribbons. As mentioned before, the reconnection flux and rates approximately balance each other. Furthermore, almost simultaneous signatures of bursts in the HXR emission suggest that the reconnection rate bursts is a real physical phenomena that could be linked with these HXR bursts, and current sheet dynamics.

\cite{Naus2022CorrelatedFlare} used high resolution IRIS observations to compare evolution of ribbon fronts with HXR emission from Fermi GBM and the Reuven Ramaty High Energy Solar Spectroscopic Imager (RHESSI; \citealt{Lin2002TheRHESSI}). They found that the ribbon front exhibits highly structured and dynamic changes in their local widths. These intermittent changes in widths are consistent with our observations of the ribbon sweeping new reconnected flux and producing bursty reconnection rates. Additionally, they found that the evolution of ribbon front widths was co-spatial and co-temporal with the bursty UV and HXR emission during the late phase of bursty reconnection of the flare. This suggests that in the early evolution of the flare, bursty reconnection does not guarantee non-thermal particle acceleration. Here we find similar decoupling between the reconnection rate and HXR emission in the early phase of the flare (see \S \ref{sec:rec_xray}).

Numerical simulations and observations have allowed the exploration of how HXR emission is suppressed by the change in the magnetic field shear of the flare loops. \cite{Qiu2022PropertiesRibbons}, found that the lag in HXR emission to the UV emission from flare ribbons corresponds to the strong to weak magnetic field shear evolution of the flare loops (inferred from flare ribbons), consistent with the 3D MHD simulation by \cite{Dahlin2022VariabilityFlares}. They presented a clear connection between the shear inferred from the flare loops and ribbons and the variation of the guide field (magnetic field component perpendicular to the reconnection plane) in the reconnection sheet, where these values cannot be measured. Other numerical experiments by \cite{Arnold2021ElectronReconnection} show that the presence of a strong guide field suppresses the particle acceleration efficiency of plasmoids.   

Figure \ref{fig:xray_rec_ind} compares the temporal evolution of the reconnection rates with the HXR emission from Fermi GBM and GOES XRS emission rates (proxy for HXRs). We found that bursts in the reconnection rates (shown with gray vertical lines in Figure \ref{fig:xray_rec_ind}) are associated with nearly co-temporal HXR emission. These results suggest that there might be some current sheet dynamics that links these two processes. Similarly, \cite{Clarke2021Quasi-periodicFlare} found co-temporal QPPs during the 2015-11-04 M3.7 flare in multiple wavelengths: 2.5 MHz, $171$ \AA{}, $1600$ \AA{}, $1-8$ \AA, and $25-50$ keV. The 2 minute periodicity found in the EUV, SXR and HXR implies a common progenitor for the QPPs. Additionally, they find that the EUV and X-ray pulsations are located in the source of HXR flare loop footpoints. Therefore, intermittently-accelerated electrons during the reconnection process form bursty emission in the EUV and X-ray. A possible explanation for this intermittent acceleration of the electrons is intermittent magnetic reconnection modulated by the PI. Our result provide additional evidence of this intermittent magnetic reconnection. We observe $2.5$ minute reconnection rate QPPs during this M3.7 flare, which provides additional evidence that oscillatory reconnection, perhaps modulated by the PI, drive the bursty emission in the HXR and SXR rates.  

In the PI driven magnetic reconnection scenario the current sheet would be filled with plasmoids that can interact and merge accelerating populations of particles that can later produce the HXR and SXR emission\citep{Drake2006ElectronReconnection,Guidoni2016MAGNETIC-ISLANDFLARES,Guidoni2022SpectralIslands}. Single magnetic island accelerators have been found to provide insufficient energy gains to reproduce HXR emission \citep{Drake2006ElectronReconnection, Guidoni2016MAGNETIC-ISLANDFLARES}. Yet,  \cite{Guidoni2022SpectralIslands} analytically showed that transport of electrons between consecutive magnetic islands with similar accelerator efficiencies leads to the non-thermal electron distribution with sufficient energy to produce HXR emission due to electron collisions in the chromosphere. Thus, the ejected plasmoid populations can merge with one another to form larger plasmoids (\citealt{Loureiro2007InstabilityChains,Uzdensky2010FastRegime}), all while accelerating particles due to magnetic mirroring within them. These large plasmoids can then reconnect with the flare loops and transfer high energy particle populations to them, which could later propagate to the chromosphere. Once in the denser chromospheric layers the HXR and SXR emissions can form as described in the standard flare model.	 

Additionally, we have found that bursts in the reconnection rates behave like QPPs, and estimated the periods for $73$ flares using the Wavelet transform. We found reconnection rate oscillation to have periods ranging from almost a minute to four minutes, with an average period of $2.5$ minutes. These oscillations appear to be related to the QPPs detections in the Fermi GBM and GOES XRS observations, even when the X-ray QPPs shown in Figures \ref{fig:fermi_rdb_periods} and \ref{fig:goes_rdb_periods} have shorter periods, on average of 1.5 minutes. We suggest that the sub-minute discrepancies are related to the lower cadence in the AIA observations (24 seconds) used to calculate the reconnection flux and rates. Features in the reconnection rate time series with a timescale lower than 24 seconds are temporally averaged out in each read by the AIA instrument. As a consequence, the time and amplitude of each of the reconnection rate bursts (local maxima in the time series) have an uncertainty. In addition, the reconnection rate wavelet transform excludes many possible sub-minute oscillations due to the 48 second Nyquist limit. Both of these consequences of the lower cadence of AIA data can lead to errors in our estimate of the reconnection rate oscillation periods. Another possibility is that the particle acceleration evident as X-ray QPP is caused by different current sheet dynamics, leading to the differences between the reconnection rate oscillations and the X-ray QPPs.  

Other mechanisms to explain QPPs in the reconnection rate include recent findings by \cite{Thurgood2017} where magnetic reconnection in a 3D null point can occur periodically and excite MHD waves. A 2D simulation showed that oscillating plasmoids produced by the subsequent merging of two initial plasmoids, can excite oscillations of the current sheet loop tops as they reconnect \citep{Jelinek2017OscillationsSheet}. \cite{Takahashi2017Quasi-periodicReconnection} used a 2D simulation of reconnection in a current sheet below an eruptive CME and found that for large values of the Lundquist number $S = 5.6 \times 10^3$, and $2.8 \times 10^4$ (PI unstable), periodic reconnection would produce reconnection jets that excite oscillations at the top and bottom of the current sheet. These jets then could form shocks that are capable of accelerating particles with similar QPP periods of the reconnection rate oscillations. We can summarize these mechanism as MHD wave driven magnetic reconnection. The coupling of these waves with the termination shock above the flare loop can accelerate the coronal particles to produce the observed X-ray QPPs.   

Our study does not address the specific contribution of each mechanism in the production of the QPPs in the reconnection rates and the HXR emission, and how the QPPs differ slightly in their periods. The spatiotemporal analysis of the flare ribbons suggests that the current sheet undergoes PI and that plasmoids and their dynamics should be related to the generation of the QPPs and bursty signatures presented in this work. Still, we are unable to determine whether the HXR emission is driven by termination shocks produced by the reconnection jets or merging of plasmoids populated with accelerated electrons. There is a probability that both mechanisms are present during magnetic reconnection in large flares. In the future, it will be of much interest to evaluate the relative contributions from each mechanism into production of QPPs in the X-ray.

Finally, we find that the delay time between the reconnection rate oscillations and the X-ray QPPs is within the range of seconds to minutes. On average both the Fermi GBM and GOES XRS QPPs are
delayed by $24$ seconds. We verify that the delay, which matches the AIA temporal cadence, is not a consequence of the numerical differentiation scheme. Specifically, we found that statistical properties of the delay distribution are independent of the finite difference scheme, when using forwards or central differentiation, to calculate the reconnection rates. When using backwards differentiation, the average delay between the QPPs decreases to 5 seconds. Yet, the overall relationship of the reconnection rate oscillation onset occurring a few seconds to minutes before the X-Ray QPPs remain the same. Therefore, the delay is either a consequence of the differences in cadences between the instrument or a physical mechanism causing the delay. \cite{Miklenic2007ReconnectionFlare,Veronig2015MagneticFlare} found similar results when comparing reconnection rates with the HXR emission observed by RHESSI and GOES SXR emission rate. They interpret the newly brightened H$_\alpha$/EUV/UV flaring kernels which are used to track the reconnection flux evolution and HXR emission to be produced by two types of non-thermal electrons. The first type would correspond to a smaller population of non-thermal electrons accelerated within the reconnection site that produces the bright H$_\alpha$/EUV/UV kernels. The second larger population of non-thermal particles would be accelerated at a different location, possibly at the flare loop tops, providing a delay in the emission burst relative to the reconnection rate burst. To our knowledge there is no observational study, nor simulation that provides evidence to this delay between flare kernels brightening, and the HXR emission and their association with these different non-thermal electron populations. Exploring the reason(s) for this delay will be the focus of a future study.	 

We hope that multi-viewing point observations will help determine which mechanism is driving the QPPs in the X-rays and reconnection rates. For example, \cite{Hayes2019} presented on-limb evidence of EUV and SXR QPPs during the X8.2 flare on 2017-09-10 with periods of approximately 65 and 150 seconds during the impulsive and decay phases. A key result from their study was that using AIA $131$ \AA{} they found downward plasma motions along the current sheet that impact the top of the flaring loop arcade. These downward motions are co-temporal with the decay phase QPPs, suggesting a link between these downward motions in the current sheet with the QPP signals. These structures could be large scale plasmoids exiting the current sheet and merging with the flare loop tops. Therefore co-temporal observation of the on-disk observations of the flare ribbon and off-limb observation of the loop structures during the flare could allow a more complete understanding of the full mechanisms at play for the generation of the QPPs in the HXR emission and reconnection rates. Coordinated observations between SDO (or other ground-based/geosynchronous instrument) and Solar Orbiter \citep{Muller2013SolarOrbiter,Muller2020TheMission} will provide opportunities for these types of studies when the two spacecrafts have a separation angle of $\approx 90^{\circ}$.

%%%%%%%%%%%%%%%%%%%%%%%%%%%%%%%%%%%%%%%%%%%%%%
\section{Conclusions}\label{sec:con}

In this study we have analyzed oscillations in the magnetic reconnection rate and compared them with QPPs in Fermi GBM and GOES XRS observations for $73$ flares from the \verb+RibbonDB+ database \citep{Kazachenko2017}. Our analysis includes the spatiotemporal comparison of flare ribbons observed with AIA $1600$ \AA{} and IRIS SJI $1330$ \AA{} and $1400$ \AA{} (when available), time series of the reconnection fluxes/rates, HXR emission ranging from 0 to 300 keV, and SXR emission/rates in the $1-8$ \AA{} and $0.5-4$ \AA{}. Our findings are summarized below:
\begin{itemize}
	\item For the first time we find that magnetic reconnection, as described by reconnection rates derived from flare ribbons, occurs in bursts. These bursts on average account for $75$\% of the total reconnection flux of each of the $73$ flares in our sample. The periods are on the order of minutes ranging from 1 to 4 minutes, and an average of 1.5 minutes.
	\item From the high-resolution IRIS SJI observations of 8 C- to X-class flares, we find that the fine structure in the flare ribbons is associated with periodic modulations in the ribbon fronts \citep{Naus2022CorrelatedFlare}, wave-break patterns, and swirls \citep{Brannon2015SPECTROSCOPICWAVES,Parker2017ModelingShear}. These fine structure patterns have been previously associated with the Plasmoid Instability, and the existence of plasmoid structures in the current sheet above the flaring loops.
	\item We find X-ray QPPs that are delayed by up to a minute from bursts in the reconnection rates. The delay time of the Fermi and GOES observations is 24 seconds on average, suggesting that the particle acceleration of non-thermal particles that interact with the chromosphere and produce the flaring ribbon kernels and the emission of SXR and HXR might happen at different locations within the current sheet.
	\item We suggest that nearly co-temporal QPP bursts in the reconnection rates and HXR emission provide evidence of oscillatory process in the reconnection region and plasmoid dynamics \citep{Kliem2000SolarReconnection,Lynch2016,Takahashi2017Quasi-periodicReconnection}. This in turn leads to bursty particle acceleration via plasmoid dynamics or wave particle interactions \citep{Takahashi2017Quasi-periodicReconnection,Guidoni2016MAGNETIC-ISLANDFLARES,Drake2006ElectronReconnection}. Our observational estimates could be used for comparison with models of bursty or oscillatory reconnection.
\end{itemize}

We acknowledge support from NASA LWS 80NSSC19K0910, 80NSSC19K0070, NASA ECIP NNH18ZDA001N, and NSF CAREER SPVKK1RC2MZ3 (MDK, MFCA). Support for this work is provided by the National Science Foundation through the DKIST Ambassadors program, administered by the National Solar Observatory and the Association of Universities for Research in Astronomy, Inc. BJL acknowledges support from NASA LWS 80NSSC21K1325, NASA XRP 80NSSC22K0674, and NSF AGS 2147399. MFCA would also like to thank Dr. Ryan French for the valuable discussions on the implementation of wavelet transform analysis to detect QPPs. 

\appendix

\section{Supplemental Figures}\label{sec:sup_mat}
\begin{figure}[!ht]
	\centering
	\includegraphics[width=\textwidth,trim={0cm 0cm 0cm 0cm}, clip]{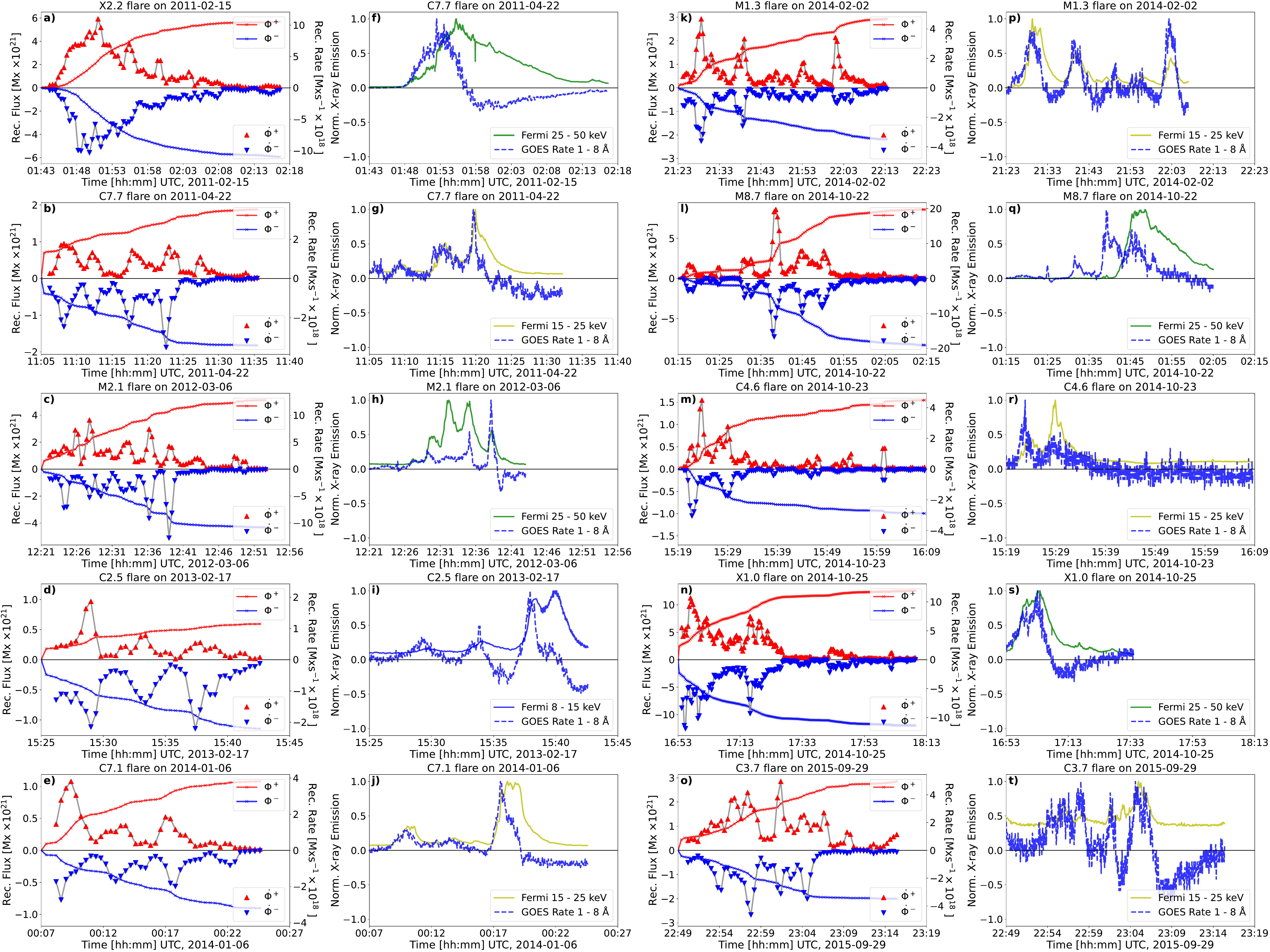}
	\caption{10 example flares with oscillations (QPPs) in the reconnection rates and X-ray lightcurves. Panels \textbf{a}-\textbf{e} and \textbf{k}-\textbf{o} show reconnection fluxes (crosses) and reconnection flux rates (triangles) where red and blue colors correspond to positive and negative polarities respectively. Panels \textbf{f}-\textbf{j} and \textbf{p}-\textbf{t} show Fermi HXR and GOES SXR rates lightcurves.}
	\label{fig:rdb_exp}
\end{figure}

Figure \ref{fig:rdb_exp} shows 10 example flares from the 73 included in the study, that exhibit clear reconnection rate oscillation on each polarity flare ribbon. Each panel shows the reconnection flux and rates derived from flare ribbons from each polarity of the active region.

\startlongtable
\begin{deluxetable}{lcccc}
	\tabletypesize{\footnotesize}
	\tablecaption{Magnetic reconnection rate oscillation period for each of the 73 flares included in our study.\label{tab:apdx}}
	\tablehead{
\colhead{Date}&\colhead{Flare start time [UTC]}&\colhead{GOES FLare Class}&\colhead{NOAA AR}&\colhead{$\mathcal{P}(\dot{\Phi})$[min.]}}
	\startdata
	2010-06-12&00:30&M2.0&11081&1.80\\
	2010-11-11&15:54&C4.3&11123&2.14\\
	2010-11-15&07:28&C2.3&11124&4.29\\
	2011-02-13&17:28&M6.6&11158&4.29\\
	2011-02-15&01:44&X2.2&11158&3.93\\
	2011-03-09&23:13&X1.5&11166&2.78\\
	2011-03-10&03:50&C2.9&11166&2.78\\
	2011-03-25&23:08&M1.0&11176&1.27\\
	2011-04-22&04:35&M1.8&11195&3.31\\
	2011-04-22&11:07&C7.7&11195&4.68\\
	2011-09-06&22:12&X2.1&11283&0.83\\
	2011-09-07&22:32&X1.8&11283&2.14\\
	2011-10-15&13:56&C5.0&11319&2.78\\
	2011-11-06&14:30&C5.3&11339&5.56\\
	2011-11-07&03:05&C3.5&11339&2.55\\
	2011-12-26&20:12&M2.3&11387&2.34\\
	2012-01-14&03:19&C2.1&11396&4.29\\
	2012-03-06&12:23&M2.1&11429&3.03\\
	2012-03-19&21:55&C3.5&11434&0.83\\
	2012-03-25&00:15&C3.0&11444&1.97\\
	2012-05-11&09:37&C2.3&11476&3.03\\
	2012-05-13&07:21&C7.0&11476&6.61\\
	2012-06-05&20:49&C4.2&11499&1.28\\
	2012-07-02&14:32&C1.7&11513&3.93\\
	2012-07-02&19:34&C2.5&11515&2.34\\
	2012-07-04&01:30&C5.1&11515&2.34\\
	2012-07-04&04:28&M2.3&11515&1.97\\
	2012-07-04&09:47&M5.3&11515&3.03\\
	2012-07-05&20:09&M1.6&11515&3.03\\
	2012-07-10&12:08&C5.5&11520&2.34\\
	2012-10-24&09:06&C2.9&11598&3.30\\
	2013-02-17&15:26&C2.5&11675&2.14\\
	2013-04-11&06:55&M6.5&11719&2.34\\
	2013-08-31&17:20&C2.6&11836&4.68\\
	2013-10-11&22:56&C6.3&11861&0.83\\
	2013-10-13&00:12&M1.7&11865&1.39\\
	2013-10-14&12:56&C8.0&11865&2.14\\
	2013-10-22&04:12&C4.0&11875&1.65\\
	2013-10-28&11:32&M1.4&11877&2.34\\
	2013-11-07&14:15&M2.4&11890&0.83\\
	2013-11-10&05:08&X1.1&11890&2.55\\
	2013-12-14&11:00&C2.3&11917&1.52\\
	2014-01-06&00:08&C7.1&11944&4.29\\
	2014-01-07&04:40&C2.4&11944&1.39\\
	2014-02-02&21:24&M1.3&11967&5.10\\
	2014-02-13&12:33&C3.0&11974&1.80\\
	2014-02-14&02:40&M2.3&11974&3.03\\
	2014-04-23&00:48&C4.3&12038&1.97\\
	2014-05-03&05:31&C5.3&12051&3.03\\
	2014-06-18&03:12&C4.0&12087&7.86\\
	2014-07-30&16:00&C9.0&12127&2.14\\
	2014-09-10&17:21&X1.6&12158&3.93\\
	2014-10-20&16:00&M4.5&12192&4.29\\
	2014-10-22&01:16&M8.7&12192&6.06\\
	2014-10-22&14:02&X1.6&12192&0.83\\
	2014-10-23&15:20&C4.6&12192&5.56\\
	2014-10-25&16:55&X1.0&12192&3.60\\
	2014-10-26&10:04&X2.0&12192&5.56\\
	2014-12-09&09:58&C8.6&12230&3.03\\
	2014-12-17&21:02&C7.1&12242&1.80\\
	2014-12-19&09:31&M1.3&12242&2.55\\
	2014-12-20&00:11&X1.8&12242&2.34\\
	2015-01-06&05:16&C6.0&12253&1.80\\
	2015-03-12&11:38&M1.6&12297&3.31\\
	2015-03-15&11:31&C6.8&12297&1.80\\
	2015-04-18&18:09&C2.9&12321&1.27\\
	2015-06-22&17:39&M6.5&12371&5.10\\
	2015-08-21&09:34&M1.4&12403&1.52\\
	2015-08-22&09:09&C5.1&12403&2.34\\
	2015-09-27&10:20&M1.9&12422&1.80\\
	2015-09-28&07:27&M1.1&12422&1.39\\
	2015-09-29&22:50&C3.7&12422&1.80\\
	2015-11-04&13:31&M3.7&12443&2.55
	\enddata
\end{deluxetable}

%%%%%%%%%%%%%%%%%%%%%%%%%%%%%%%%%%%%%%%%%%%%%%
% Bibliography
%%%%%%%%%%%%%%%%%%%%%%%%%%%%%%%%%%%%%%%%%%%%%%
\bibliographystyle{aasjournal}
\bibliography{RecPaper}

\end{document}